\documentclass[
aps,
prx,
twocolumn,
superscriptaddress,
floatfix,
longbibliography,
nofootinbib
]{revtex4-2}

\usepackage{graphicx}
\usepackage{dcolumn}
\usepackage{bm}
\usepackage{amsmath}
\usepackage{amssymb}
\usepackage{txfonts}
\usepackage{color}
\usepackage{xcolor}
\usepackage{listings}
\usepackage{cprotect}
\usepackage{bbm}
\usepackage{multirow}
\usepackage{braket}

\usepackage{hyperref}
\usepackage{pdfpages}
\usepackage{pgffor}

\newcommand{\im}{\mathrm{i}}
\def\avg#1{\hat{#1}_{\rm avg}}

\makeatletter
\AtBeginDocument{\let\LS@rot\@undefined}
\makeatother

\graphicspath{{./figures/}}

\hypersetup{
  colorlinks=true,
  linkcolor=[rgb]{0.60,0.00,0.00},
  citecolor=[rgb]{0.00,0.00,0.60},
  urlcolor=[rgb]{0.00,0.00,0.60},
  setpagesize=false
}

\begin{document}
\title{
Noise-stabilized discrete time crystals on digital quantum processors
}
\author{Kazuya Shinjo}
\affiliation{Computational Quantum Matter Research Team, RIKEN Center for Emergent Matter Science (CEMS), Wako, Saitama 351-0198, Japan}
\author{Kazuhiro Seki}
\affiliation{Quantum Computational Science Research Team, RIKEN Center for Quantum Computing (RQC), Wako, Saitama 351-0198, Japan}
\author{Seiji Yunoki}
\affiliation{Computational Quantum Matter Research Team, RIKEN Center for Emergent Matter Science (CEMS), Wako, Saitama 351-0198, Japan}
\affiliation{Quantum Computational Science Research Team,
RIKEN Center for Quantum Computing (RQC), Wako, Saitama 351-0198, Japan}
\affiliation{Computational Materials Science Research Team,
RIKEN Center for Computational Science (R-CCS), Kobe, Hyogo 650-0047, Japan}
\affiliation{Computational Condensed Matter Physics Laboratory, 
RIKEN Pioneering Research Institute (PRI), Saitama 351-0198, Japan}

\date{\today}


\begin{abstract}
Floquet many-body phases such as discrete time crystals (DTCs) are typically fragile to imperfections, and stabilizing them on noisy quantum hardware remains a central challenge in nonequilibrium quantum physics. 
Here, we use IBM Eagle and Heron superconducting processors to implement Floquet dynamics of a kicked Ising model on two-dimensional Kagome lattices, engineered via ancilla-assisted embeddings into the heavy-hex connectivity of the devices. 
By combining error-mitigated measurements on quantum hardware with matrix-product-state simulations incorporating an ancilla-noise model constructed from experimental device data, we observe long-lived subharmonic magnetization oscillations that are stabilized---rather than destroyed---by structured quantum noise. 
Across different two-dimensional lattice geometries, increasing cases beyond Kagome lattices, and with or without boundary symmetry-charge pumping, ancilla errors effectively act as spatiotemporal disorder that induces stochastic sign flips of the Ising couplings, providing a unified mechanism for robust period-doubling responses. 
When symmetry-charge pumping is present, intrinsic boundary-localized $\pi$ modes cooperate with this disorder to yield a boundary-assisted DTC characterized by suppressed scrambling and sharply localized dynamics. 
In contrast, in implementations without pumping, the noiseless dynamics rapidly thermalize and exhibit no subharmonic order, whereas the same noise process alone generates a DTC-like long-lived subharmonic response over experimentally accessible time windows. 
These results identify engineered ancilla noise as a practical control knob for inducing, stabilizing, and geometrically tailoring nonequilibrium dynamical order on scalable superconducting quantum processors.
\end{abstract}

\maketitle

\section{Introduction}

Understanding nonequilibrium dynamics in interacting quantum many-body systems is a central challenge in modern physics and a key avenue for demonstrating the scientific utility of programmable quantum processors.
While classical tensor-network methods efficiently capture low-entanglement regimes~\cite{Orus2019,Weimer2021,Cirac2021,Tindall23b,Patra2024}, the rapid growth of entanglement, particularly in higher spatial dimensions, severely limits access to late-time dynamics, motivating direct experimental probes of real-time unitary evolution on quantum hardware.

Driven many-body systems provide a particularly rich setting, exemplified by discrete time crystals (DTCs)~\cite{Sacha2018,Khemani2019,Else2020,Guo2020,Sacha2020,Zaletel2023}, which exhibit robust subharmonic responses that break the discrete time-translation symmetry imposed by periodic driving. 
While disorder-driven many-body localization can stabilize DTC order in one dimension~\cite{Else2016,Khemani2016,vonKeyserlingk2016,Yao2017,Zhang2017,Ippoliti2021,Mi2022,Frey2022,Zhang2022}, such mechanisms are widely expected to be unstable in higher dimensions. 
This has motivated disorder-free routes to DTCs, including prethermal regimes~\cite{Else2017,Machado2020,Kyprianidis2021,Beatrez2023,Jin2025,Jiang2025} and clean, strongly interacting regimes~\cite{Huang2018,Pizzi2021,Santini2022,Collura2022,Huang2023,Shinjo2024}.
In most quantum-hardware realizations, however, noise is treated purely as detrimental, washing out subharmonic signatures before the relevant time scales can be reached. 
Throughout this work, we use the term ``DTC'' to denote robust, long-lived subharmonic responses that persist over many drive periods in experimentally accessible system sizes, often in a prethermal regime, rather than an idealized infinite-time ordered phase in the strict thermodynamic limit.

Current superconducting quantum processors face an additional bottleneck beyond noise: fixed device connectivity restricts the direct realization of nontrivial lattice geometries, such as frustrated lattices or lattices with higher coordination. 
Embedding flexible lattice structures into such fixed connectivity, together with quantitative noise control, broadens the range of nonequilibrium dynamical phases accessible on near-term hardware and enables systematic studies of how noise and geometry jointly shape driven quantum dynamics.

\begin{figure*}[htbp]
\includegraphics[width=1.0\textwidth]{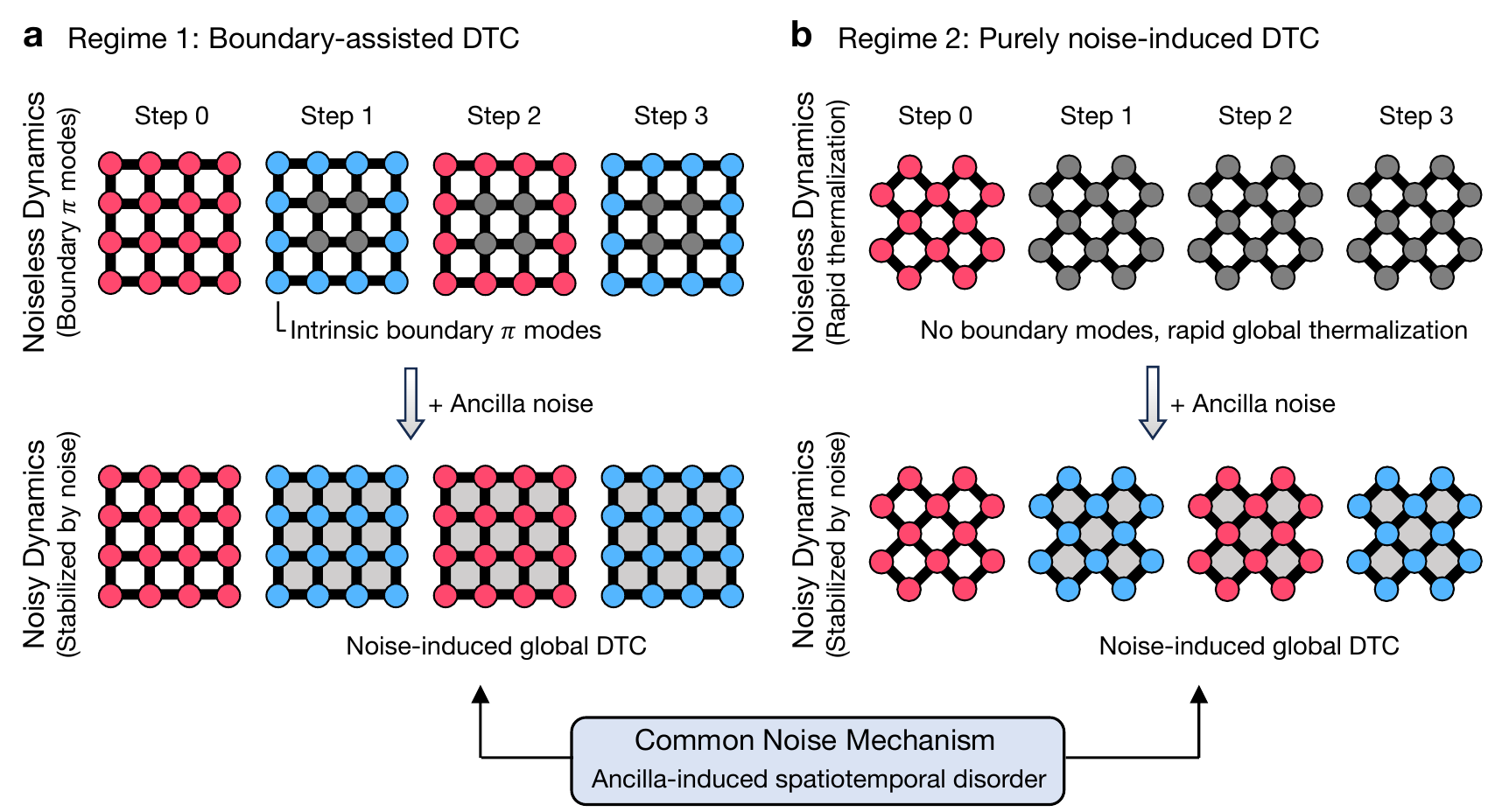}
\caption{
    \textbf{Noise stabilization of subharmonic dynamics in complementary regimes.}
    Schematic overview of the two noise-assisted regimes studied in this work, unified by a common microscopic mechanism in which ancilla-induced spatiotemporal disorder manifests as stochastic sign flips of effective Ising couplings. 
    Colored sites (denoted by circles) indicate the stroboscopic local magnetization over Floquet steps (blue/red: opposite subharmonic states; gray: thermalized).
    The background shading indicates the presence or absence of ancilla noise (white: noise-free; light gray: noise-affected)
    \textbf{a}, Boundary-assisted (boundary$+$noise) regime. In the noiseless limit, the dynamics support boundary symmetry-charge pumping and host localized boundary $\pi$ modes, while the bulk rapidly thermalizes. In the presence of ancilla noise, stochastic sign flips of effective Ising coupling cooperate with the boundary $\pi$ modes at the pumped sites to suppress scrambling, yielding a sharply localized subharmonic response across the lattice. 
    \textbf{b}, Noise-only regime, in which no pumping sites are present. The noiseless dynamics rapidly scramble and thermalize, whereas the same ancilla noise slows thermalization and stabilizes long-lived subharmonic oscillations across the lattice.
}
\label{fig:overview}
\end{figure*}

Here we report noise-assisted DTC dynamics on digital quantum processors implementing two-dimensional Kagome lattices, with additional results on two-dimensional Lieb lattices. 
Using coordination-three qubits as ancillas, we embed these geometries into the heavy-hex connectivity of two generations of IBM superconducting devices, Eagle and Heron, with the latter featuring improved gate fidelities (see Supplementary Information, Sec.~S1.6 and Table~S3). 
On these embedded lattices, we implement a periodically driven kicked Ising model with a transverse field~\cite{Pineda2014}, initialize the system in a product state, and probe the resulting dynamics via local magnetization measurements. 
To elucidate the origin of the observed subharmonic responses, error-mitigated measurements on quantum hardware are compared with classical matrix-product-state (MPS) simulations incorporating ancilla-noise models constructed from experimental data, which quantitatively reproduce the observed magnetization dynamics over up to 40 Floquet cycles.
Figure~\ref{fig:overview} summarizes the central physical picture and the two complementary noise-assisted regimes explored in this work.

Within this effective noise-induced description, we identify a unified mechanism that yields two complementary regimes, determined by whether the underlying noiseless Floquet dynamics support boundary pumping. 
In one regime, the noiseless dynamics generate symmetry-charge pumping at selected boundary sites and host intrinsic boundary-localized $\pi$ modes with robust period-doubling oscillations. 
Ancilla-induced spatiotemporal disorder, which effectively manifests as stochastic sign flips of the Ising couplings, does not simply degrade these modes; instead, it cooperates with the boundary $\pi$ modes to stabilize a boundary-assisted DTC, in which scrambling is blocked at the pumped sites and the subharmonic response remains sharply localized across the lattice.

In the complementary regime, realized in Floquet implementations without symmetry-charge pumping such that boundary $\pi$ modes are absent in the noiseless limit, the same ancilla-noise process gives rise to a long-lived DTC response. 
In the presence of ancilla noise, stochastic sign flips of the effective Ising interactions slow thermalization and stabilize subharmonic magnetization oscillations across the lattice. 
By contrast, removing this noise causes the corresponding noiseless dynamics to rapidly scramble and thermalize into a highly entangled state with no trace of subharmonic order. 
This pumping-free regime therefore demonstrates that ancilla noise alone can induce and stabilize DTC behavior. 

More broadly, these results establish digital quantum processors as a versatile platform for exploring nonequilibrium dynamical order, in which engineered quantum noise functions not merely as an experimental limitation but as a tunable ingredient that induces, stabilizes, and geometrically tailors subharmonic responses. 
Our work highlights engineered ancilla noise 
as a key physical resource for realizing and controlling nonequilibrium quantum matter, and broadens the landscape of dynamical phases accessible on near-term quantum hardware.


\section{Results}
\label{sec:results}

\subsection{Floquet circuit implementation on digital quantum hardware}
\label{sec:floquet_impl}

We study a kicked Ising model with $L$ system qubits, governed by a time-periodic Hamiltonian $\hat H(t)=\hat H(t+T)$ with period $T$,
\begin{equation}\label{eq-hamiltonian}
\hat{H}(t)=
\begin{cases}
h_{x}\sum_{i=0}^{L-1}\hat{X}_{i}, & 0\le t<T/2,\\[2pt]
-J\sum_{\langle i,j\rangle}\hat{Z}_{i}\hat{Z}_{j}, & T/2\le t<T,
\end{cases}
\end{equation}
where $\hat{X}_i$ and $\hat{Z}_i$ denote Pauli operators acting on qubit $i$, and $\langle i,j\rangle$ labels nearest-neighbor pairs on the target lattice. 
The corresponding single-cycle Floquet unitary is given by 
\begin{equation}\label{eq:uf}
\hat{U}_{\rm F} (\theta_x)
=\left[\prod_{\langle i,j\rangle}\hat{R}_{Z_iZ_j}(\theta_J)\right]
 \left[\prod_{i}\hat{R}_{X_i}(\theta_x)\right],
\end{equation}
with $\hat{R}_{Z_iZ_j}(\theta_J)=\exp \left[-\im \theta_J\hat{Z}_i\hat{Z}_j/2 \right]$ and 
$\hat{R}_{X_i}(\theta_x)=\exp \left[-\im \theta_x\hat{X}_i/2\right]$.
Throughout this work, we fix $\theta_J=-\pi/2$, a parameter choice known to support robust subharmonic responses in related Floquet settings~\cite{Shinjo2024}. 
To quantify deviations from perfect period doubling, we introduce a perturbation parameter $\epsilon=\pi-\theta_x$ and focus on the weakly perturbed regime $0<\epsilon\ll\pi$.

We realize these Floquet dynamics on IBM digital quantum processors using ancilla-assisted circuits. 
System qubits encode spins on two-dimensional Kagome lattices, while ancilla qubits mediate the required two-qubit couplings within the native heavy-hex connectivity of the devices. 
Additional results for Lieb lattices are presented in the Supplementary Information (Sec.~S4.1). 
The single-period Floquet unitary $\hat U_{\rm F}(\theta_x)$ is compiled into native single- and two-qubit gates together with ancilla-mediated mid-circuit operations, enabling flexible control over both the drive parameters and the effective ancilla-noise processes realized on the hardware.

The system is initialized in a fully polarized product state in the $Z$ basis, and the stroboscopic dynamics at times $t=nT$ are given by 
$|\psi(t)\rangle=(\hat{U}_{\rm F})^{n}|\psi(0)\rangle$.
As a primary observable, we consider the local magnetization
$\langle \hat{Z}_j(t)\rangle = \langle\psi(t)|\hat{Z}_j|\psi(t)\rangle$, which can equivalently be expressed in the Heisenberg picture as $\hat Z_j(t)=(\hat U_{\rm F}^\dagger)^n \hat Z_j (\hat U_{\rm F})^n$, with the initial state $|\psi(0)\rangle$ defined at $t=0$. 
At $\epsilon=0$, a single Floquet period flips the sign of each $\hat{Z}_j$, producing perfect period-doubling oscillations. 
Our central question is how robust this subharmonic response remains for small detuning $\epsilon>0$ in the presence of realistic hardware noise.

\begin{figure*}[htbp]
\includegraphics[width=1.0\textwidth]{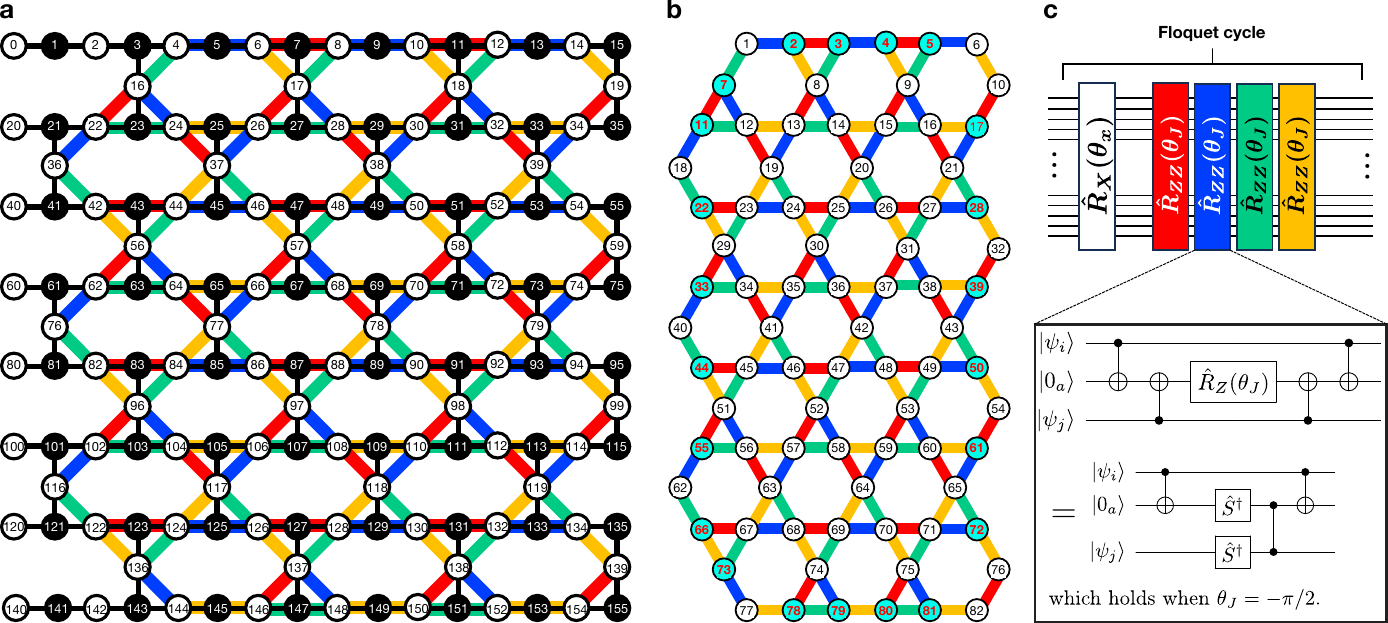}
\caption{
    \textbf{Two-qubit gate connectivity and geometry of the Kagome82 lattice.}
    \textbf{a}, Kagome82 lattice constructed on the heavy-hex architecture of the \texttt{ibm\_marrakesh} device, comprising 156 qubits in total. 
    White and black circles denote system qubits ($|i\rangle$ and $|j\rangle$) and ancilla qubits ($|a\rangle$), located at sites with coordination numbers two and three, respectively, on the heavy-hex architecture. 
    The four layers of $\hat{R}_{Z_{i}Z_{j}}(\theta_{J})$ gates applied within a single Floquet cycle are indicated in red, blue, green, and yellow. 
    \textbf{b}, Geometry of the Kagome82 lattice showing only the system qubits, renumbered to define a one-dimensional path for matrix-product-state (MPS) construction.
    Green circles indicate system qubits with coordination number three in the Kagome82 lattice. 
    \textbf{c}, Schematic representation of the single-cycle Floquet operator $\hat{U}_{\rm F}$.
    Red, blue, green, and yellow boxes correspond to the four layers of $\hat{R}_{Z_{i}Z_{j}}(\theta_{J})$ gates, applied in parallel as indicated in panel~\textbf{a}.
    White boxes represent the product of $\hat{R}_{X_i}$ gates. 
    Horizontal lines denote qubits on which the gates act. 
    Each $\hat{R}_{Z_{i}Z_{j}}(\theta_{J})$ gate acting on system qubits  $i$ and $j$ is implemented using ancilla-assisted constructions: either four CNOT gates together with a single-qubit $\hat{R}_{Z_{a}}(\theta_{J})$ gate acting on an ancilla qubit $|0_{a}\rangle$, or, for $\theta_J=-\pi/2$, three CNOT and CZ gates together with two phase gates $\hat S$.
}
\label{fig:geometry_Kagome82}
\end{figure*}

In our experiments, we use one Eagle device (\texttt{ibm\_kyiv}) and two Heron devices (\texttt{ibm\_torino} and \texttt{ibm\_marrakesh}) from IBM. 
By selecting qubits with coordination number two on the heavy-hex lattice as system qubits and adjacent coordination-three qubits as ancillas, we realize large-scale Kagome lattice instances, including an 82-qubit instance (Kagome82) that serves as our primary example. 
Figure~\ref{fig:geometry_Kagome82} illustrates the construction of Kagome82 on \texttt{ibm\_marrakesh}. 
Additional experiments on other 53-qubit Kagome implementations (Kagome53-I and Kagome53-II), as well as on a 40-qubit Lieb lattice (Lieb40), are presented later in this section and summarized in Supplementary Table~S1.
Further details of the ancilla-assisted embeddings into the heavy-hex architecture are provided in Supplementary Sec.~S1.1.

Within this experimental platform, different choices of lattice geometry and boundary termination naturally give rise to two complementary dynamical regimes, distinguished by whether the corresponding noiseless dynamics exhibit boundary symmetry-charge pumping and an associated boundary-localized $\pi$ mode, as diagnosed below via boundary magnetization. 
Implementations supporting such a boundary mode exhibit subharmonic responses stabilized jointly by the boundary $\pi$ mode and ancilla-induced noise; we refer to this as a boundary-assisted DTC (``boundary$+$noise'' regime). 
In contrast, implementations that lack any boundary $\pi$ mode can still exhibit long-lived subharmonic oscillations, even though such oscillations are entirely absent in the corresponding noiseless dynamics. 
In this case, the oscillations are stabilized purely by ancilla-induced noise, demonstrating a noise-only mechanism for sustaining subharmonic response without any boundary pumping. 
Representative examples of both regimes are discussed below.

On these lattices, each system qubit interacts with at most four neighbors, such that the two-qubit gates $\hat{R}_{Z_iZ_j}(\theta_J)$ can be applied in four parallel layers, as indicated by the colored bonds in Fig.~\ref{fig:geometry_Kagome82}a. 
Because interacting system qubits are not directly connected in the heavy-hex architecture, each $\hat{R}_{Z_iZ_j}(\theta_J)$ gate is implemented via an ancilla qubit $a$ adjacent to both qubits $i$ and $j$, using a three-qubit phase gadget~\cite{Cowtan2020}, 
\begin{equation}\label{eq:RZZ0}
\exp[-\im \theta_J \hat{Z}_a\hat{Z}_i\hat{Z}_j/2]\,|0_a\rangle\otimes|\psi\rangle
=|0_a\rangle\otimes\hat{R}_{Z_iZ_j}(\theta_J)|\psi\rangle,
\end{equation}
where the ancilla ideally returns to its initial state $|0_a\rangle$. 
In practice, decomposing this gadget into native two-qubit gates introduces noise on both the system and ancilla qubits. 
As shown in Sec.~\ref{sec:NoisyModel}, noise accumulated on the ancillas can be captured by an effective noise model. 
Within this picture, ancilla-induced spatiotemporal disorder plays a central role in slowing thermalization and stabilizing the observed subharmonic responses in both the ``boundary$+$noise'' and ``noise-only'' regimes. 
With this ancilla-assisted Floquet architecture in place, we now turn to the measured magnetization dynamics and their comparison with noisy MPS simulations.

\subsection{Boundary-assisted regime on Kagome82}
\label{sec:noise-dtc-kagome82}

We probe DTC behavior by measuring the stroboscopic magnetization after repeated applications of the Floquet unitary. 
As our primary observable, we consider the magnetization averaged over all system qubits in the set $A$ with $|A|=L$, 
\begin{align}\label{eq:Zav}
\langle\avg{Z}(t)\rangle=\frac{1}{|A|}\sum_{j\in A}\langle\hat Z_{j}(t)\rangle.
\end{align}
The expectation values are evaluated using the error-mitigation protocol described in Methods (Sec.~\ref{sec:errormitigation}), which renormalizes the raw signal by the trivial $\theta_x=\pi$ reference to compensate for global depolarization of the system qubits. 
On the Kagome82 lattice realized on \texttt{ibm\_marrakesh}, we observe a robust subharmonic response at half the drive frequency, characteristic of a long-lived DTC, over a broad range of transverse-field angles $\theta_x$.

\begin{figure}[htbp]
\includegraphics[width=0.47\textwidth]{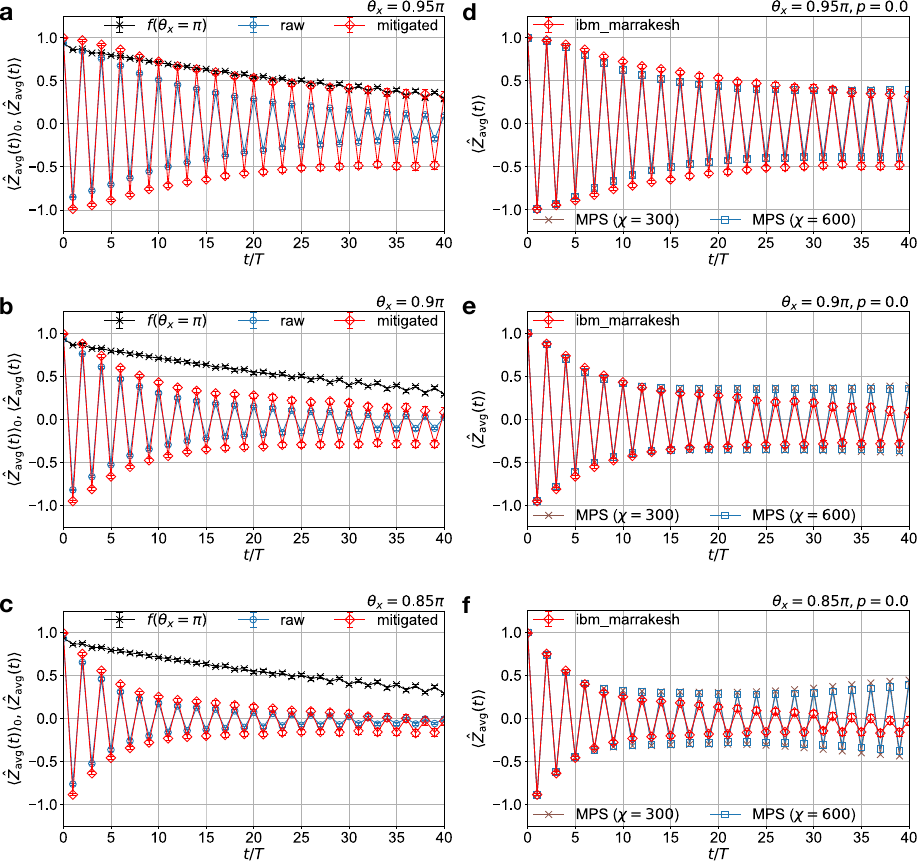}
\caption{
    \textbf{Error-mitigated magnetization dynamics on the Kagome82 lattice.}
    The Kagome82 lattice is realized on the \texttt{ibm\_marrakesh} device.
    \textbf{a}--\textbf{c}, Raw experimental magnetization data $\langle \avg{Z}(t) \rangle_{0}$ (blue circles), together with the corresponding renormalization factor $f(\theta_x=\pi)=|\langle \avg{Z}(t) \rangle_{0,\theta_x=\pi}|$ (black crosses). 
    Error-mitigated results obtained using Eq.~(\ref{eq-norm}) are shown as red diamonds.
    \textbf{d}--\textbf{f}, Comparison of error-mitigated experimental magnetization data (red diamonds) with noiseless ($p=0$) MPS simulations using bond dimensions $\chi=300$ (brown crosses) and $\chi=600$ (blue squares). 
    The transverse-field parameters are $\theta_{x}=0.95\pi$ in panels~\textbf{a} and~\textbf{d}, $\theta_{x}=0.9\pi$ in panels~\textbf{b} and~\textbf{e}, and $\theta_{x}=0.85\pi$ in panels~\textbf{c} and~\textbf{f}. 
    Each $\hat{R}_{Z_i Z_j}$ gate is implemented using $M_\mathrm{CNOT}=3$ native two-qubit gates. 
    }
\label{fig:localZ_Kagome82}
\end{figure}

Figure~\ref{fig:localZ_Kagome82}a--c shows the raw magnetization data $\langle \avg{Z}(t)\rangle_0$ for $\theta_{x}=0.95\pi$, $0.9\pi$, and $0.85\pi$, respectively, together with the corresponding renormalization factor $f(\theta_x=\pi)=|\langle \avg{Z}(t)\rangle_{0,\theta_x=\pi}|$. 
The raw data (blue circles) display clear period-doubling oscillations up to $40$ Floquet cycles, although their amplitude decays due to accumulated quantum noise. 
Applying the renormalization protocol defined in Eq.~(\ref{eq-norm}) restores the oscillation amplitude, yielding the error-mitigated signal (red diamonds). 
For the Kagome82 implementation, each $\hat R_{Z_i Z_j}$ gate is compiled into $M_{\rm CNOT}=3$ native two-qubit gates (Fig.~\ref{fig:geometry_Kagome82}c), such that a single Floquet period already involves several hundred two-qubit gates. 
By $t/T=40$, the total circuit volume $\nu$, defined as the cumulative number of native two-qubit gates, reaches $v = 17,040$, exceeding the volume $v=15,000$ used in previous experiments on the heavy-hex lattice of \texttt{ibm\_torino}, where clean DTCs were observed up to $t/T=100$~\cite{Shinjo2024} (see Supplementary Information, Sec.~S1.2, for gate-count details).

To assess the reliability of the observed oscillations, we compare the error-mitigated experimental data with noiseless MPS simulations, shown in Fig.~\ref{fig:localZ_Kagome82}d--f. 
At early times ($t/T\lesssim 10$), the experimental results and noiseless MPS simulations are in good agreement. 
At later times, however, noticeable deviations emerge, particularly for $\theta_{x}=0.9\pi$ and $0.85\pi$, reflecting the influence of noise components beyond those captured by a purely global depolarizing model acting on the system qubits.

\begin{figure}[htbp]
\includegraphics[width=0.47\textwidth]{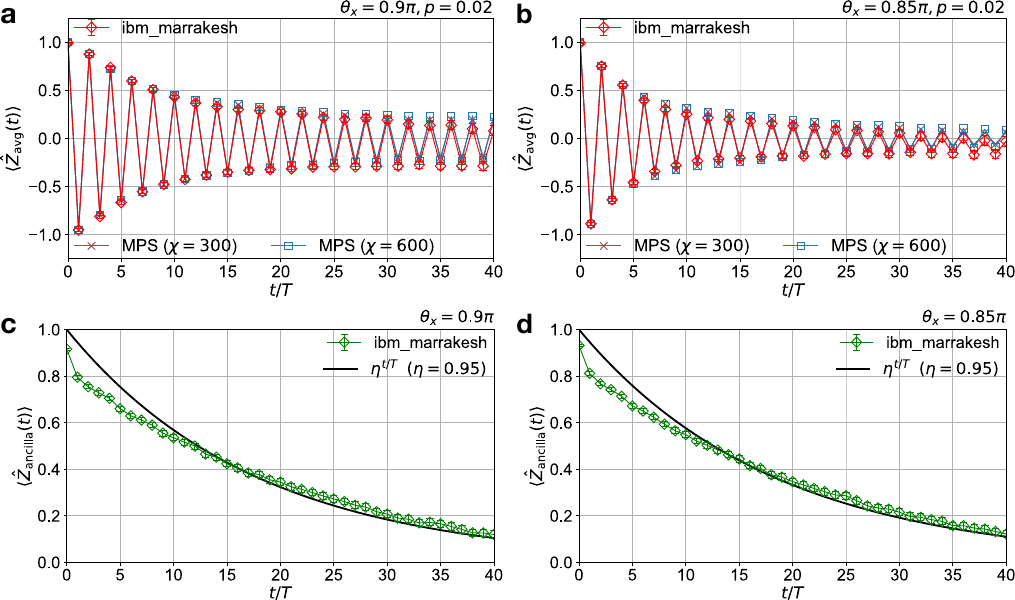}
\caption{
    \textbf{Effect of ancilla noise on magnetization dynamics on the Kagome82 lattice.} 
    The Kagome82 lattice is realized on the \texttt{ibm\_marrakesh} device. 
    \textbf{a,b}, Comparison of error-mitigated experimental magnetization data (red diamonds) with noisy MPS simulations at noise parameter $p=0.02$, using bond dimensions $\chi=300$ (brown crosses) and $\chi=600$ (blue squares). 
    \textbf{c,d}, Magnetization averaged over all ancilla qubits, $\langle \hat Z_\text{ancilla}(t) \rangle$ (green diamonds), together with a single-exponential fit $\eta^{t/T}$ (black line). These are raw experimental data without error mitigation. 
    The transverse-field parameters are $\theta_{x}=0.9\pi$ in panels~\textbf{a} and~\textbf{c}, and $\theta_{x}=0.85\pi$ in panels~\textbf{b} and~\textbf{d}.
}
\label{fig:localZ_Kagome82noise}
\end{figure}

To incorporate ancilla-induced noise, we perform MPS simulations based on the effective noise model described in Methods (Secs.~\ref{sec:ancillanoise} and \ref{sec:NoisyModel}), in which ancilla errors are represented as stochastic sign flips of the interaction angle $\theta_J$ with probability $p$ per Floquet cycle. 
The value of $p$ is identified from the ancilla fidelity via Eq.~(\ref{eq:p-q}). 
Figure~\ref{fig:localZ_Kagome82noise}a,b compares the error-mitigated experimental data with noisy MPS simulations at $p=0.02$, showing substantially improved agreement relative to the noiseless cases shown in Fig.~\ref{fig:localZ_Kagome82}e,f. 
The same ancilla noise also manifests directly in the ancilla magnetization $\langle \hat Z_\text{ancilla}(t)\rangle$, defined as the raw magnetization averaged over all ancilla qubits, which ideally remains at unity but decays rapidly in the experiment (see Fig.~\ref{fig:localZ_Kagome82noise}c,d). 
Fitting $\langle \hat Z_\text{ancilla}(t)\rangle$ to a single-exponential form $\eta^{t/T}$ yields an average ancilla fidelity $\eta\simeq 0.95$. 
Equation~(\ref{eq:p-q}) then gives a corresponding sign-flip probability $p=(1-\eta)/2\approx 0.02$, consistent with the value used in the noisy MPS simulations (see also Sec.~\ref{sec:NoisyModel} and Supplementary Information Sec.~S2.2).

\begin{figure*}[btp]
\includegraphics[width=1.0\textwidth]{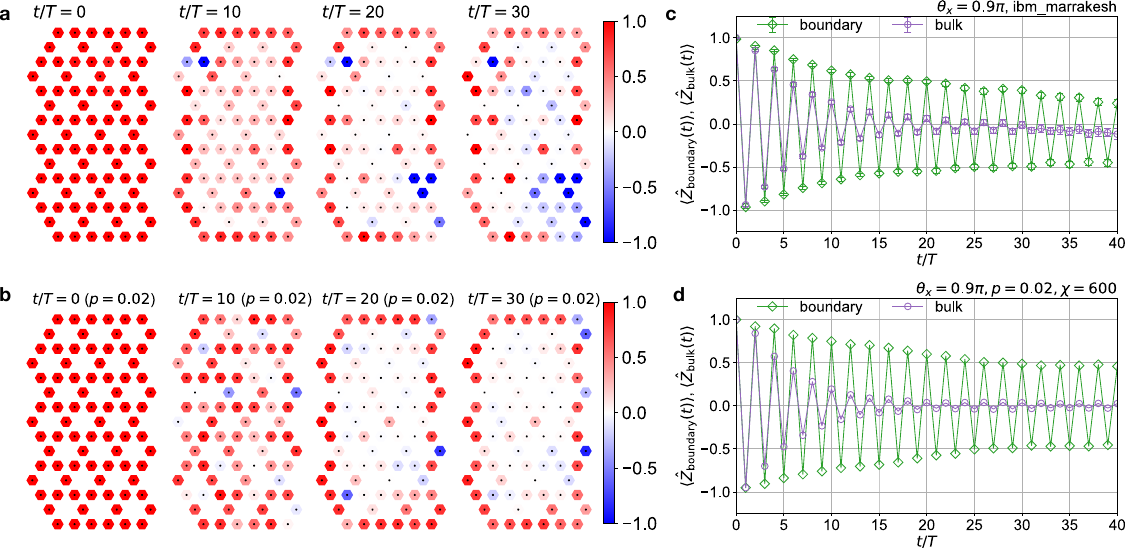}
\caption{
    \textbf{Time evolution of magnetization in the kicked Ising model on the Kagome82 lattice.}
    \textbf{a}, Snapshots of the error-mitigated local magnetization $\langle \hat Z_{j}(t) \rangle$ at $t/T=0$, $10$, $20$, and $30$, measured on \texttt{ibm\_marrakesh}.
    \textbf{b}, Corresponding results obtained from noisy MPS simulations with $p=0.02$ and bond dimension $\chi=600$. 
    \textbf{c}, Error-mitigated magnetization averaged over the boundary qubits, $\langle \hat Z_\text{boundary}(t) \rangle$ (green diamonds), and over the bulk qubits, $\langle \hat Z_\text{bulk}(t) \rangle$ (purple circles), measured on \texttt{ibm\_marrakesh}.
    The full set of boundary sites is specified in Supplementary Information Sec.~S1.4. 
    The bulk includes all system sites except those located at the boundary. 
    \textbf{d}, Same as panel~\textbf{c}, but obtained from noisy MPS simulations with $p=0.02$ and $\chi=600$. 
    The transverse-field parameter is set to $\theta_{x}=0.9\pi$.
}
\label{fig:Zdist_Kagome82}
\end{figure*}

The spatial structure of the period-doubling oscillations on Kagome82 is shown in Fig.~\ref{fig:Zdist_Kagome82}. 
Figure~\ref{fig:Zdist_Kagome82}a displays the error-mitigated local magnetizations, $\langle \hat{Z}_{j}(t)\rangle=\langle \hat{Z}_{j}(t)\rangle_{0}/\langle \hat{Z}_\text{avg}(t)\rangle_{0,\theta_{x}=\pi}$, at selected times, revealing that the largest oscillation amplitudes are localized near the boundary. 
Noisy MPS simulations with $p=0.02$ (Fig.~\ref{fig:Zdist_Kagome82}b) exhibit a closely similar spatial pattern. 
To quantify this behavior, we separate the average magnetization into boundary and bulk contributions: Fig.~\ref{fig:Zdist_Kagome82}c,d compare the boundary-averaged and bulk-averaged signals for the experiment and noisy MPS simulations, respectively. 
Here, the bulk includes all system sites except those located at the boundary. 
In the noiseless limit, the contrast between boundary and bulk becomes even more pronounced (see Sec.~\ref{sec:symmetry_pumping}). 
A small subset of boundary sites exhibits reduced oscillation amplitudes due to local connectivity and spatiotemporal disorder arising from ancilla-induced noise, as analyzed in detail in Sec.~\ref{sec:symmetry_pumping} and Supplementary Information Sec.~S4.

\begin{figure}[btp]
\includegraphics[width=0.47\textwidth]{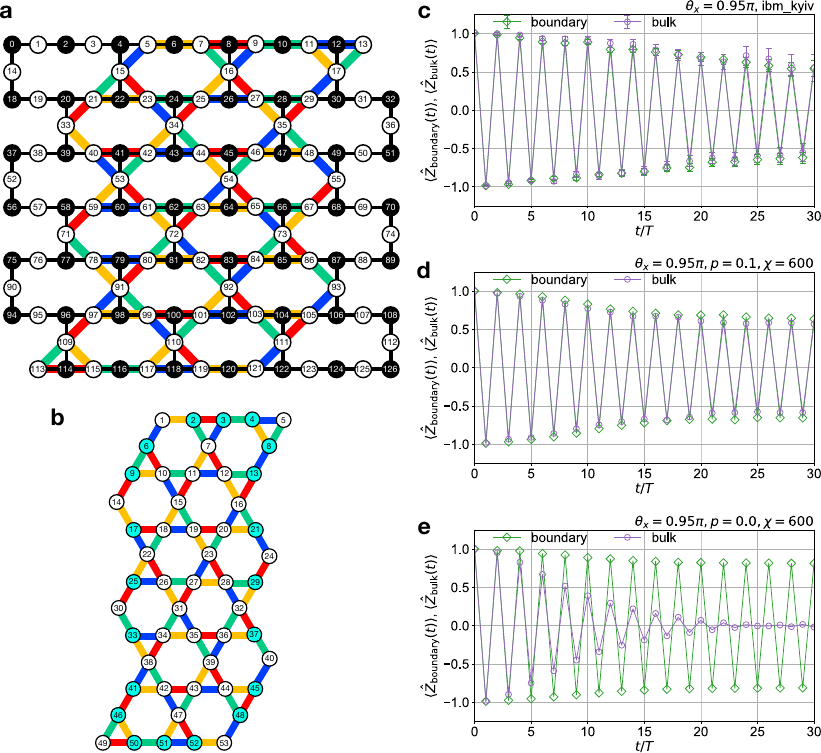}
\caption{
    \textbf{Time evolution of magnetization in the kicked Ising model on the Kagome53-I lattice.}
    \textbf{a}, Kagome53-I lattice constructed on the heavy-hex architecture of the \texttt{ibm\_kyiv} device with $127$ qubits. White and black circles denote system and ancilla qubits, located at sites with coordination numbers two and three, respectively. The four layers of $\hat{R}_{Z_{i}Z_{j}}(\theta_J)$ gates applied within a single Floquet cycle are indicated in red, blue, green, and yellow. 
    \textbf{b}, Geometry of the Kagome53-I lattice showing only the system qubits, with renumbered labels defining a one-dimensional path for MPS construction. Green circles indicate system qubits with coordination number three in the Kagome53-I lattice, and boundary qubits are located along the outer edges (see Supplementary Information Sec.~S1.4 for the full list).
    \textbf{c}, Error-mitigated magnetization averaged over the boundary qubits, $\langle \hat Z_\text{boundary}(t) \rangle$ (green diamonds), and over the bulk qubits, $\langle \hat Z_\text{bulk}(t) \rangle$ (purple circles), measured on \texttt{ibm\_kyiv}.
    \textbf{d}, Same quantities as in panel~\textbf{c}, but obtained from noisy MPS simulations with $p=0.1$ and bond dimension $\chi=600$. 
    \textbf{e}, Same as panel~\textbf{d}, but obtained from noiseless MPS simulations with $\chi=600$.
    The transverse-field parameter is set to $\theta_{x}=0.95\pi$.
}
\label{fig:localZedge095pi_Kagome53noCorner_kyiv}
\end{figure}

For Kagome82 at low ancilla noise, the most pronounced period-doubling oscillations are confined to the boundary, and the emergence of a truly bulk DTC in the thermodynamic limit cannot be definitively established from the system sizes studied here. 
As the ancilla noise level increases, however, the intrinsic boundary $\pi$ mode cooperates with ancilla-induced spatiotemporal disorder to induce subharmonic oscillations on bulk sites, yielding a boundary-assisted DTC in which a localized boundary response coexists with an extended bulk subharmonic component. 
To explore this regime, we construct a 53-qubit Kagome lattice (Kagome53-I) on \texttt{ibm\_kyiv}, using coordination-three sites on the heavy-hex graph as ancillas (see Fig.~\ref{fig:localZedge095pi_Kagome53noCorner_kyiv}a,b). 
Figure~\ref{fig:localZedge095pi_Kagome53noCorner_kyiv}c shows that on \texttt{ibm\_kyiv}, which exhibits higher noise levels than \texttt{ibm\_marrakesh}, period-doubling oscillations appear with comparable amplitude in both the boundary and bulk regions.

Noisy MPS simulations with ancilla-noise parameter $p=(1-\eta)/2\simeq 0.1$ reproduce the experimental magnetization oscillations on Kagome53-I (Fig.~\ref{fig:localZedge095pi_Kagome53noCorner_kyiv}d), where the ancilla fidelity $\eta\simeq 0.82$ is extracted from direct measurements of the ancilla magnetization (see Fig.~\ref{fig:localZ095pi_Kagome53withCorner}g). 
In contrast, in the absence of ancilla noise ($p=0$), the subharmonic response remains strongly localized at the boundary (Fig.~\ref{fig:localZedge095pi_Kagome53noCorner_kyiv}e), consistent with boundary-localized $\pi$ modes in the underlying Floquet spectrum discussed below.

These observations on the Kagome82 and Kagome53-I lattices indicate that the subharmonic response is organized by boundary physics. 
To isolate this mechanism and make connect it with the spatial structure observed in the data, we next characterize the noiseless dynamics in terms of symmetry-charge pumping, which gives rise to boundary-localized $\pi$ modes and an information-blockade mechanism. 
This noiseless characterization also provides a reference point for the following sections, where we examine how subharmonic responses persist in implementations without symmetry-charge pumping and to what extent ancilla-induced noise alone can sustain such dynamics.

\subsection{Symmetry-charge pumping and information blockade}
\label{sec:symmetry_pumping}

Motivated by the boundary-localized oscillations observed on the Kagome lattices, we now analyze the microscopic structure of the underlying noiseless Floquet dynamics. 
For $\theta_J=-\pi/2$, the kicked Ising Floquet circuit admits a description in terms of symmetry-charge pumping: the doubled Floquet evolution $(\hat U_{\rm F})^2$ acts differently on a subset of sites $j\in P$ with odd coordination number (see Supplementary Information Sec.~S3). 
We refer to qubits in this set $P$ as charge-pumped sites. 
In all lattices exhibiting a boundary mode, the set $P$ lies along the boundary and is determined by the local connectivity and lattice termination; these sites are highlighted by green circles in Figs.~\ref{fig:geometry_Kagome82}b and \ref{fig:localZedge095pi_Kagome53noCorner_kyiv}b.
We find the same pumping-based structure on Lieb and other lattice geometries; additional results are provided in Supplementary Information Secs.~S4.1 and S4.3.

A compact way to express the pumping structure is through the local action of the doubled Floquet operator on Pauli-$X$ operators (see Supplementary Information Sec.~S3):
\begin{equation}
[\hat U_\text{F}(\pi-\epsilon)]^{-2}\,\hat X_j\,[\hat U_\text{F}(\pi-\epsilon)]^{2} \simeq \omega_j\,\hat X_j,
\quad
\omega_j=
\begin{cases}
-1,& j\in P,\\
+1,& j\notin P,
\end{cases}
\label{eq:omega_main}
\end{equation}
when $\epsilon$ is small. 
Equation~\eqref{eq:omega_main} implies the existence of $\pi$-paired Floquet doublets~\cite{Khemani2019,Else2020} localized near the pumped region, yielding boundary-localized $\pi$ modes and long-lived period-doubling oscillations of the local magnetization in the noiseless dynamics.

\begin{figure}[htbp]
\includegraphics[width=0.47\textwidth]{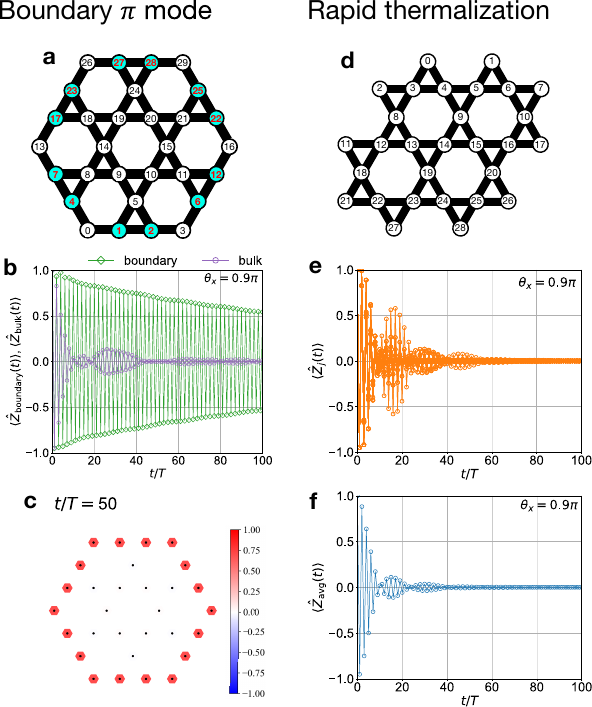}
\caption{
\textbf{Noiseless dynamics in pumped and pumping-free Kagome geometries.}
\textbf{a--c}, Results for the pumped Kagome lattice ($P\neq\varnothing$; Kagome30). 
\textbf{a}, Lattice geometry; green circles denote charge-pumped sites $j\in P$ (which, in this geometry, coincide with coordination-three sites). 
\textbf{b}, Boundary-averaged magnetization, $\langle Z_\text{boundary}(t)\rangle$ (green diamonds), and bulk-averaged magnetization, $\langle Z_\text{bulk}(t)\rangle$ (purple circles). 
\textbf{c}, Snapshot of the local magnetization $\langle \hat Z_j(t)\rangle$ at $t/T=50$. 
\textbf{d--f}, Results for the pumping-free Kagome lattice ($P=\varnothing$; Kagome29). 
\textbf{d}, Lattice geometry. 
\textbf{e}, Site-resolved local magnetization $\langle \hat Z_j(t)\rangle$ for all system sites. 
\textbf{f}, Magnetization averaged over all system sites, $\langle Z_\text{avg}(t)\rangle$. 
The transverse-field parameter is set to $\theta_{x}=0.9\pi$, and all results are obtained from noiseless statevector simulations. The definitions of the lattice geometries are summarized in Supplementary Information Sec.~S1 and Table~S2. 
}
\label{fig:localZ_statevector_Kagome}
\end{figure}

To isolate a noiseless reference, we perform exact statevector simulations on smaller Kagome lattices, namely 
Kagome30 with $P\neq\varnothing$ and Kagome29 with $P=\varnothing$ 
(Fig.~\ref{fig:localZ_statevector_Kagome}; see also Supplementary Information Sec.~S1 and Table~S2).
Figure~\ref{fig:localZ_statevector_Kagome}a--c illustrates the pumped case, where the boundary-averaged signal retains a pronounced subharmonic component, while the bulk relaxes more rapidly, consistent with the boundary-localized $\pi$-mode dynamics implied by Eq.~\eqref{eq:omega_main}.
In addition, the pumped sites suppress entanglement propagation across them, effectively isolating neighboring sites and producing an information-blockade effect that stabilizes localized oscillations. 
Related blockade phenomena have been reported in one-dimensional Floquet settings~\cite{Sur2023,Sedlmayr2023}.
Further geometry-dependent manifestations and complementary diagnostics are presented in Supplementary Information Secs.~S4.2, S4.3, and S4.4.

By contrast, for the pumping-free Kagome geometry ($P=\varnothing$), both the site-resolved local magnetization and the system-averaged magnetization decay rapidly under noiseless dynamics (Fig.~\ref{fig:localZ_statevector_Kagome}d--f), indicating that no site exhibits a long-lived subharmonic response. 
These pumping-free dynamics therefore provide a clear baseline for Sec.~\ref{sec:noise-induced}, where we show that ancilla-induced noise can sustain long-lived subharmonic responses even when $P=\varnothing$.

\subsection{Noise-induced discrete time crystals without boundary modes}\label{sec:noise-induced}

We now examine the complementary regime in which $P=\varnothing$, such that the Floquet dynamics lack boundary $\pi$ modes altogether. 
In this pumping-free setting, the central question is whether ancilla-induced quantum noise alone can stabilize DTC behavior in the absence of any symmetry-charge-pumping mechanism.

\begin{figure}[htbp]
\includegraphics[width=0.47\textwidth]{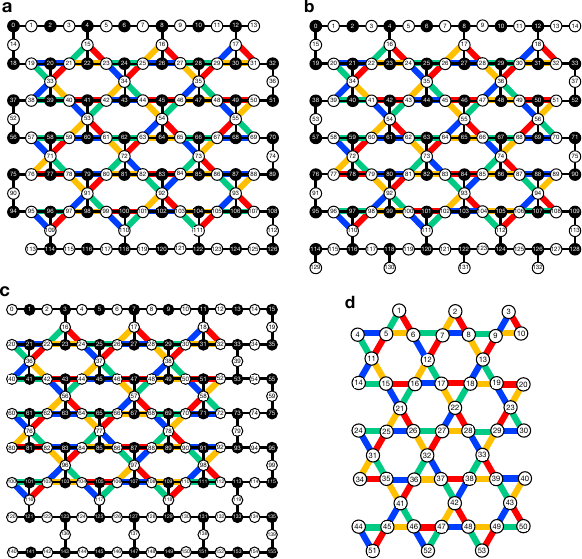}
\caption{
    \textbf{Two-qubit gate connectivity and geometry of the Kagome53-II lattice.}
    \textbf{a}, Kagome53-II lattice constructed on the heavy-hex architecture of the \texttt{ibm\_kyiv} device (comprising 127 qubits in total). 
    \textbf{b}, Same lattice embedded on \texttt{ibm\_torino} (133 qubits).
    \textbf{c}, Same lattice embedded on \texttt{ibm\_marrakesh} (156 qubits).
    In panels~\textbf{a--c}, white and black circles denote system and ancilla qubits located at sites with coordination numbers two and three, respectively, on the heavy-hex architecture; colored bonds indicate the four layers of $\hat{R}_{Z_{i}Z_{j}}(\theta_J)$ gates applied within a single Floquet cycle.
    \textbf{d}, Geometry of the Kagome53-II lattice showing only the system qubits, renumbered to define the one-dimensional path used for MPS construction. 
    In this connectivity, no system qubit has coordination number three, and the corresponding Floquet description yields $P=\varnothing$.
}
\label{fig:Kagome53withCorner}
\end{figure}

To probe this regime on hardware, we implement a Kagome lattice explicitly designed to have no charge-pumped sites. 
By appropriately selecting system qubits and ancillas within the heavy-hex connectivity, we realize the Kagome53-II lattice shown in Fig.~\ref{fig:Kagome53withCorner}d on three IBM quantum devices: \texttt{ibm\_kyiv} (Fig.~\ref{fig:Kagome53withCorner}a), \texttt{ibm\_torino} (Fig.~\ref{fig:Kagome53withCorner}b), and \texttt{ibm\_marrakesh} (Fig.~\ref{fig:Kagome53withCorner}c). 
The boundary termination of Kagome53-II differs from that of Kagome53-I studied earlier in this section; in particular, the effective Floquet description yields $P=\varnothing$ for Kagome53-II, whereas Kagome53-I hosts boundary-localized pumped sites ($P\neq\varnothing$).
Additional experiments on Kagome53-II using an alternative gate decomposition with $M_{\rm CNOT}=4$ native two-qubit gates, which further enhances the effective ancilla-noise level, are reported in Supplementary Information Sec.~S5.1.

\begin{figure*}[htbp]
\includegraphics[width=1.0\textwidth]{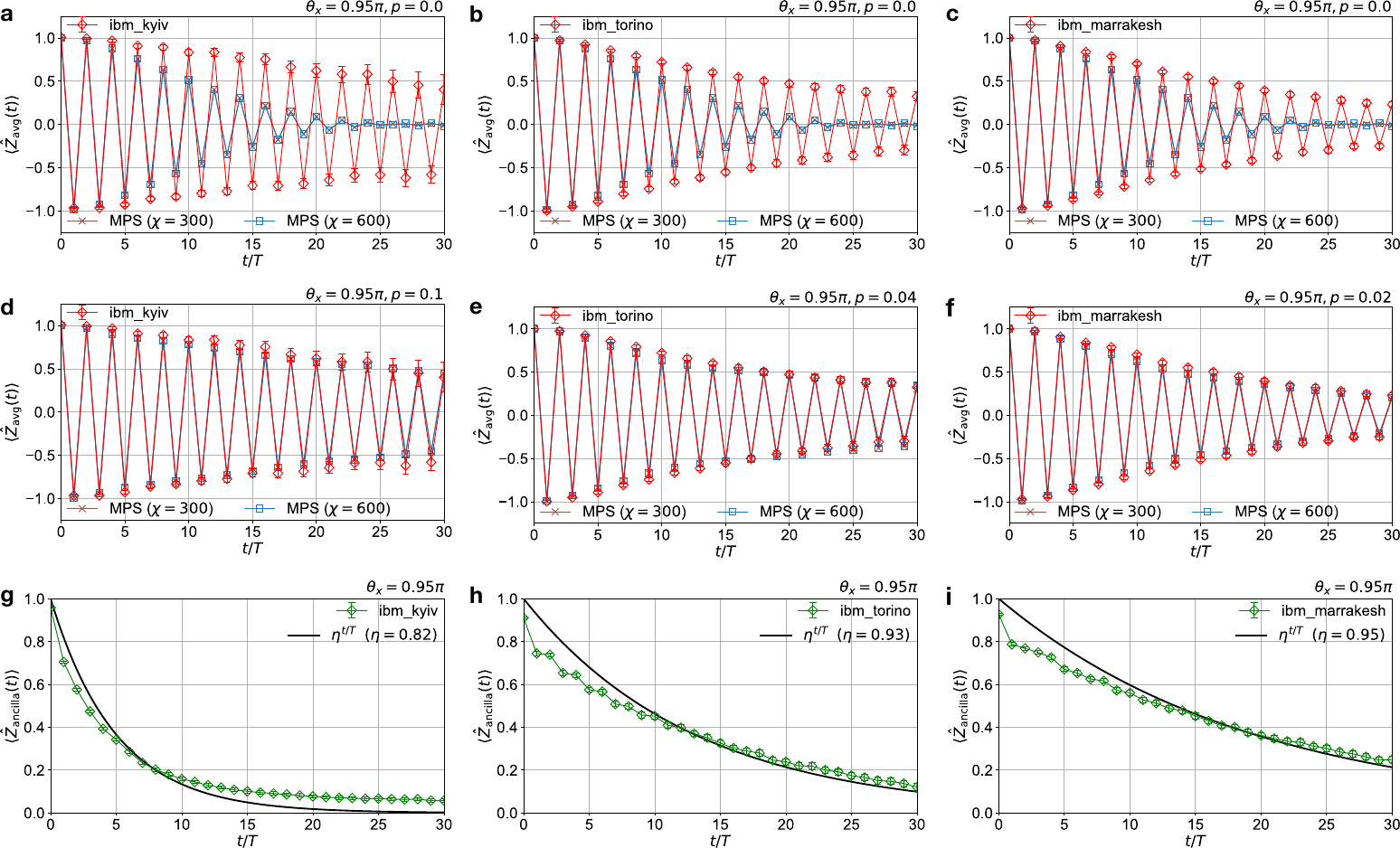}
\caption{
    \textbf{Time evolution of magnetization in the kicked Ising model on the Kagome53-II lattice.}
    The transverse-field parameter is set to $\theta_x=0.95\pi$, and the dynamics are measured on \texttt{ibm\_kyiv} (left panels), \texttt{ibm\_torino} (middle panels), and \texttt{ibm\_marrakesh} (right panels). 
    \textbf{a--c}, Error-mitigated magnetization averaged over all system qubits, $\langle \avg{Z}(t) \rangle$, obtained from the quantum devices (red diamonds), compared with noiseless MPS simulations using bond dimensions $\chi=300$ (brown crosses) and $\chi=600$ (blue squares). 
    \textbf{d--f}, Same experimental data compared with noisy MPS simulations incorporating ancilla-induced noise. 
    The noise parameters are set to $p=0.1$ in panel~\textbf{d}, $p=0.04$ in panel~\textbf{e}, and $p=0.02$ in panel~\textbf{f}, chosen to approximate $(1-\eta)/2$ for each device. 
    \textbf{g--i}, Magnetization averaged over ancilla qubits, $\langle \hat Z_\text{ancilla}(t) \rangle$ (green diamonds), together with single-exponential fits $\eta^{t/T}$ (black lines), from which the average ancilla fidelity $\eta$ is extracted. 
    These ancilla data are shown without error mitigation. 
    In all cases, each $\hat{R}_{Z_{i}Z_{j}}(\theta_J)$ gate is implemented using $M_\text{CNOT}=3$ native two-qubit gates. See Supplementary Information Fig.~S14 for results obtained using the implementation with $M_\text{CNOT}=4$ native two-qubit gates.  
}
\label{fig:localZ095pi_Kagome53withCorner}
\end{figure*}

The magnetization dynamics observed on these quantum devices are summarized in Fig.~\ref{fig:localZ095pi_Kagome53withCorner}. 
In the absence of charge-pumped sites, and hence boundary modes, period-doubling oscillations are expected to decay rapidly, as confirmed by the noiseless MPS simulations (blue squares) in Fig.~\ref{fig:localZ095pi_Kagome53withCorner}a. 
In contrast, the error-mitigated experimental data (red diamonds) in Fig.~\ref{fig:localZ095pi_Kagome53withCorner}a--c display enhanced period-doubling oscillations with markedly extended lifetimes. 
Remarkably, the Eagle device \texttt{ibm\_kyiv}, which exhibits the highest noise levels among the three processors, shows the largest subharmonic amplitudes. 
This device dependence directly implicates ancilla-induced quantum noise as the origin of the stabilized DTC response in the absence of symmetry-charge pumping.

To quantify the role of ancilla noise, we compare the experimental data with noisy MPS simulations in which ancilla errors are modeled as stochastic sign flips of the interaction angle $\theta_J$, occurring with probability $p$ per Floquet cycle. 
For each device, we choose $p$ according to Eq.~(\ref{eq:p-q}), namely $p\simeq(1-\eta)/2$, where $\eta$ is the average ancilla fidelity extracted from the decay of $\langle \hat Z_\text{ancilla}(t)\rangle$ (Fig.~\ref{fig:localZ095pi_Kagome53withCorner}g--i; see also Sec.~\ref{sec:NoisyModel}). 
As shown in Fig.~\ref{fig:localZ095pi_Kagome53withCorner}d--f, the noisy MPS simulations with $p=0.1$ (\texttt{ibm\_kyiv}), $0.04$ (\texttt{ibm\_torino}), and $0.02$ (\texttt{ibm\_marrakesh}) reproduce both the amplitude and the lifetime of the experimental subharmonic oscillations in close agreement with the hardware data. 
Among the three devices, the noise-induced DTC response is again most pronounced on \texttt{ibm\_kyiv}, which exhibits the highest two-qubit error rates. As discussed in Supplementary Information Sec.~S5.1, this feature becomes even more pronounced when the two-qubit gate $\hat R_{Z_i Z_j}(\theta_J)$ is implemented using $M_{\rm CNOT}=4$ native two-qubit gates.

From exponential fits to $\langle \hat Z_\text{ancilla}(t)\rangle$, we obtain average ancilla fidelities per Floquet cycle of $\eta\simeq 0.82$, $0.93$, and $0.95$ for \texttt{ibm\_kyiv}, \texttt{ibm\_torino}, and \texttt{ibm\_marrakesh}, respectively. 
These values are consistent with the estimates based on Eq.~(\ref{eq:eta}) using the average CNOT or CZ error rates listed in Supplementary Table~S3, providing an independent determination of $p=(1-\eta)/2$ used in the noisy MPS simulations.

These results demonstrate that ancilla-induced noise can stabilize a DTC response even in systems that lack symmetry-charge pumping and boundary-localized modes. 
In Methods (Sec.~\ref{sec:NoisyModel}), we show that the effective noise model parameterized by $p=(1-\eta)/2$ captures the crossover from rapid thermalization at $p=0$ to a noise-stabilized DTC regime at intermediate $p$ and yields good agreement with experimental data across different devices and geometries. 
Detailed noise-strength dependence and out-of-time-ordered correlator (OTOC) analyses on smaller Kagome lattices, which further elucidate the scrambling dynamics in this noise-induced regime, are presented in Supplementary Information Secs.~S5.3 and~S5.4.

\section{Discussion}
\label{sec:discussion}

We have demonstrated that IBM superconducting quantum processors can probe nonequilibrium Floquet dynamics on programmable two-dimensional Kagome and Lieb lattices engineered via ancilla-assisted embeddings into the heavy-hex architecture, with the Lieb lattice results presented in the Supplementary Information. 
By incorporating ancilla-noise effects in classical simulations through a noise model with parameters extracted from device data, we achieve quantitative agreement between error-mitigated magnetization measurements on quantum hardware and noisy MPS simulations over tens of Floquet cycles, capturing both the amplitude and the lifetime of the subharmonic response and thereby establishing a controlled framework for interpreting noisy quantum dynamics.

Within this framework, we identify a unified noise-induced mechanism that stabilizes long-lived subharmonic responses across all geometries studied. 
This mechanism manifests in two complementary regimes, depending on whether the underlying noiseless Floquet dynamics support symmetry-charge pumping. 
When symmetry-charge pumping generates intrinsic boundary-localized $\pi$ modes, ancilla-induced spatiotemporal disorder cooperates with these modes to produce a boundary-assisted DTC, characterized by suppressed scrambling and sharply localized subharmonic oscillations. 
In implementations without symmetry-charge pumping, the same noise process alone stabilizes a long-lived subharmonic response, whereas the corresponding noiseless dynamics rapidly scramble and evolve toward a highly entangled state devoid of subharmonic order.

These findings show that structured quantum noise---here introduced via ancilla qubits---can serve as a tunable ingredient for inducing, stabilizing, and tailoring nonequilibrium phases of matter on near-term quantum hardware. 
While our specific noise model is tailored to IBM's heavy-hex architecture and its corresponding embedding strategy, the resulting ancilla-induced sign-flip processes arise generically in a broad class of ancilla-mediated entangling schemes, suggesting that similar noise-engineered dynamical phases may be accessible on other platforms. 
More broadly, ancilla-assisted architectures provide a flexible route to implementing effective interactions and complex geometries beyond native hardware connectivity, thereby offering a scalable pathway toward digital quantum simulations of frustrated magnets, strongly correlated materials, and lattice gauge theories on future large-scale quantum platforms.


\section{Methods}

\subsection{Error mitigation via a global depolarizing model}
\label{sec:errormitigation}

To obtain reliable estimates of magnetization dynamics from quantum hardware, we employ an error-mitigation scheme based on a phenomenological global depolarizing channel~\cite{Swingle2018,Vovrosh2021,Urbanek2021}.  
For a set of system qubits $A$ with $|A|=L$, the average magnetization $\langle \avg{Z}(t)\rangle$ is defined in Eq.~(\ref{eq:Zav}). 
Experimentally, $\langle \avg{Z}(t)\rangle$ is obtained from projective measurements in the computational basis using $N_{\rm shots}=2^{12}$ samples per time step on Heron devices and $N_{\rm shots}=2^{14}$ on Eagle devices. Statistical uncertainties are estimated from the sample standard deviation of the measurement outcomes. No dynamical decoupling, zero-noise extrapolation, or probabilistic error cancellation is applied unless explicitly stated.

In the ideal kicked Ising model with $\theta_x=\pi$, one has  
$|\langle \avg{Z}(t)\rangle|=1$ for all $t$.  
In contrast, the corresponding raw experimental value  
$f(\theta_x=\pi):=|\langle \avg{Z}(t)\rangle_{0,\theta_x=\pi}|$  
decays in time due to noise.  
Under a global depolarizing channel applied after each Floquet cycle, the noisy expectation value of an observable $\hat O$ satisfies
\begin{align}
\langle \hat O(t)\rangle_0
=f\,\langle\hat O(t)\rangle
+(1-f)\,\frac{{\rm Tr}[\hat O]}{2^L},
\end{align}
where $f$ denotes a phenomenological depolarization parameter.  
Since ${\rm Tr}[\avg{Z}(t)]=0$ and the ideal value satisfies $|\langle \avg{Z}(t)\rangle|=1$ at $\theta_x=\pi$, we obtain  
\begin{align}
f(\theta_x=\pi)=|\langle \avg{Z}(t)\rangle_{0,\theta_x=\pi}|.
\end{align}

For general $\theta_x$, the corresponding ideal expectation value $|\langle \avg{Z}(t)\rangle|$ is not known \emph{a priori}. Here we approximate the depolarization factor as weakly dependent on $\theta_x$ and set $f(\theta_x)\approx f(\pi)$, yielding the renormalization
\begin{align}\label{eq-norm}
\langle \avg{Z}(t)\rangle
=\frac{\langle\avg{Z}(t)\rangle_0}
     {|\langle\avg{Z}(t)\rangle_{0,\theta_x=\pi}|}.
\end{align}
We refer to $\langle \avg{Z}(t)\rangle$ defined in Eq.~(\ref{eq-norm}) as the error-mitigated average magnetization; this is the quantity plotted, for example, in Figs.~\ref{fig:localZ_Kagome82} and \ref{fig:localZedge095pi_Kagome53noCorner_kyiv}.  
This protocol has been validated in previous studies of kicked Ising dynamics on heavy-hex devices~\cite{Shinjo2024,Frey2022,Mi2021} and provides reliable magnetization estimates over tens of Floquet cycles.

\subsection{Ancilla depolarization as an effective source of coherent sign-flip errors}
\label{sec:ancillanoise}

The global depolarizing model described above captures signal decay in system qubits but does not account for errors accumulated in ancilla qubits, which mediate the implementation of $\hat R_{Z_iZ_j}(\theta_J)$ gates acting on pairs of system qubits.   
Depolarizing noise acting on an ancilla initialized in $|0_a\rangle$ or $|1_a\rangle$ satisfies
\begin{align}
\mathcal{E}(|0_a\rangle\langle0_a|)
&=\frac{1+\eta_a}{2}|0_a\rangle\langle0_a|
 +\frac{1-\eta_a}{2}|1_a\rangle\langle1_a|,
\label{eq:depolarize1}\\
\mathcal{E}(|1_a\rangle\langle1_a|)
&=\frac{1+\eta_a}{2}|1_a\rangle\langle1_a|
 +\frac{1-\eta_a}{2}|0_a\rangle\langle0_a|,
\label{eq:depolarize2}
\end{align}
where $\eta_a$ denotes the effective ancilla fidelity per Floquet cycle. 
After $n=t/T$ cycles, the ancilla magnetization obeys 
\begin{align}
\langle \hat Z_a(t)\rangle=\eta_a^{\,t/T}.
\end{align}
Thus ancilla depolarization induces a stochastic bit-flip process with probability
\begin{align}\label{eq:q}
q_a=\frac{1-\eta_a}{2}.
\end{align}

The three-qubit phase gadget implementing $\hat R_{Z_iZ_j}(\theta_J)$ satisfies 
\begin{align}
\exp[-\im\theta_J\hat Z_a\hat Z_i\hat Z_j/2]\,
|0_a\rangle\otimes|\psi\rangle
&=|0_a\rangle\otimes \hat R_{Z_iZ_j}(\theta_J)|\psi\rangle,\\
\exp[-\im\theta_J\hat Z_a\hat Z_i\hat Z_j/2]\,
|1_a\rangle\otimes|\psi\rangle
&=|1_a\rangle\otimes \hat R_{Z_iZ_j}(-\theta_J)|\psi\rangle,
\label{eq:RZZ1}
\end{align}
so that an ancilla bit flip reverses the sign of $\theta_J$. 
Because the gadget does not reset the ancilla qubit, its state can fluctuate stochastically throughout the Floquet evolution, generating spatiotemporal patterns of coherent sign errors in the Ising couplings.
In Secs.~\ref{sec:noise-dtc-kagome82} and \ref{sec:noise-induced}, we extract average ancilla fidelities from experimental measurements of $\langle \hat Z_a(t)\rangle$ on Kagome82, Kagome53-I, and Kagome53-II, which determine the ancilla-error probability used in the noisy MPS simulations described below.

\subsection{Effective stochastic model for classical simulations}
\label{sec:NoisyModel}

To enable a direct comparison between experiments and classical simulations, we incorporate ancilla-induced errors into an effective stochastic model that eliminates the ancilla degrees of freedom and encodes their noisy influence as stochastic sign flips of the interaction parameter $\theta_J$.  
Time evolution is generated using either exact statevector methods or the two-site time-dependent variational principle~\cite{Haegeman2011} with bond dimension $\chi$. The same simulation settings are used for both noiseless and noisy MPS simulations shown in Figs.~\ref{fig:localZ_Kagome82noise}, \ref{fig:Zdist_Kagome82}, \ref{fig:localZedge095pi_Kagome53noCorner_kyiv}, and \ref{fig:localZ095pi_Kagome53withCorner}.

For each ancilla $a$, we introduce a binary stochastic variable $\xi_a\in\{0,1\}$, initialized to $\xi_a=0$.  
Each time a gate $\hat R_{Z_iZ_j}$ associated with ancilla $a$ is applied, $\xi_a$ flips with probability $p$,
\begin{align}\label{eq:xi}
\xi_a \leftarrow 1-\xi_a 
\quad \text{with probability } p,
\end{align}
and the effective system evolution replaces 
\begin{align}\label{eq:noisyRZZ}
\hat R_{Z_iZ_j}(\theta_J)
\;\longmapsto\;
\hat R_{Z_iZ_j}\bigl([1-2\xi_a]\theta_J\bigr),
\end{align}
so that $\xi_a=1$ corresponds to the sign reversal $\theta_J\mapsto -\theta_J$.  
In terms of the Hamiltonian in Eq.~(\ref{eq-hamiltonian}), this is equivalent to implementing a dynamically evolving disorder pattern via  
$-J\hat Z_i\hat Z_j \to -J[1-2\xi_a]\hat Z_i\hat Z_j$.

The probability $p$ is constrained by experimental data. 
Noisy classical simulations incorporating ancilla-noise effects reproduce the experimentally observed magnetization dynamics of the system qubits most accurately when
\begin{align}\label{eq:p-q}
p\simeq q_a=\frac{1-\eta}{2},
\end{align}
where $\eta$ denotes the average ancilla fidelity per Floquet cycle, extracted from measurements of $\langle \hat Z_a(t)\rangle$ or estimated from device-reported two-qubit gate error rates (CNOT or CZ). 
If an ancilla participates in $M_a$ phase gadgets per Floquet cycle and each gadget is compiled into $M_{\rm CNOT}$ native two-qubit gates, then
\begin{align}\label{eq:eta}
\eta \simeq (1-\varepsilon_{\rm CNOT})^{M_aM_{\rm CNOT}}
\simeq 1-M_aM_{\rm CNOT}\varepsilon_{\rm CNOT},
\end{align}
where $\varepsilon_{\rm CNOT}$ denotes the average two-qubit gate error rate obtained from the device calibration data.  
Typical values in our Kagome embeddings are $M_a\simeq 3$, while Lieb embeddings yield $M_a\simeq 2$.  
Validation of Eqs.~(\ref{eq:p-q}) and (\ref{eq:eta}) is provided in Sec.~\ref{sec:noise-induced} and Supplementary Information Sec.~S2.1.

We refer to MPS simulations that incorporate the stochastic updates in Eqs.~(\ref{eq:xi}) and (\ref{eq:noisyRZZ}) into the Floquet dynamics governed by $\hat U_{\rm F}(\theta_x)$ in Eq.~(\ref{eq:uf}) as noisy MPS simulations. The one-dimensional paths adopted for MPS construction in different lattice geometries are indicated in the corresponding figures, for example, Fig.~\ref{fig:geometry_Kagome82}b for Kagome82, Fig.~\ref{fig:localZedge095pi_Kagome53noCorner_kyiv}b for Kagome53-I, and Fig.~\ref{fig:Kagome53withCorner}d for Kagome53-II. 
As shown in Secs.~\ref{sec:noise-dtc-kagome82} and \ref{sec:noise-induced}, this stochastic sign-flip model accurately reproduces the experimentally observed slowing of thermalization and the stabilization of subharmonic oscillations across both the boundary-assisted (``boundary$+$noise'') and noise-induced (``noise-only'') DTC regimes. 
In this sense, classical simulations provide a quantitatively controlled framework for interpreting the experimental dynamics, highlighting a complementary interplay between quantum hardware experiments and classical computation rather than a fundamental dichotomy.


\begin{acknowledgments}
We are grateful to the IBM Quantum team for technical support, in particular Atsushi Matsuo and Toru Imai. 
We also benefited from valuable discussions with Netanel Lindner, Orli Alberton, and Eyal Bairey at Qedma.   
This work was supported in part by the New Energy and Industrial Technology Development Organization (NEDO), Japan (Project No. JPNP20017). 
We acknowledge support by the Japan Society for the Promotion of Science (JSPS), KAKENHI (Grant Nos.
JP21H04446, 
JP22K03520, and
JP23K13066) 
from the Ministry of Education, Culture, Sports, Science and Technology (MEXT), Japan. 
We also thank the Japan Science and Technology Agency (JST) for support through COI-NEXT (Grant No. JPMJPF2221) and MEXT for the Program for Promoting Research of the Supercomputer Fugaku (Grant No. MXP1020230411).  
Further support was provided by the UTokyo Quantum Initiative, 
the RIKEN TRIP initiative (RIKEN Quantum and Many-Body Electron Systems), and the Center of Excellence (COE) Research Grant in Computational Science from Hyogo Prefecture and Kobe City through the Foundation for Computational Science.
Part of the numerical simulations was carried out on the HOKUSAI supercomputer at RIKEN and the supercomputer system at the D3 center, Osaka University, through the HPCI System Research Project (Project ID: hp250062).
The MPS simulations were performed using the ITensor library~\cite{Fishman2022}. 
\end{acknowledgments}

\bibliography{bibdtc}

\begin{thebibliography}{42}%
\makeatletter
\providecommand \@ifxundefined [1]{%
 \@ifx{#1\undefined}
}%
\providecommand \@ifnum [1]{%
 \ifnum #1\expandafter \@firstoftwo
 \else \expandafter \@secondoftwo
 \fi
}%
\providecommand \@ifx [1]{%
 \ifx #1\expandafter \@firstoftwo
 \else \expandafter \@secondoftwo
 \fi
}%
\providecommand \natexlab [1]{#1}%
\providecommand \enquote  [1]{``#1''}%
\providecommand \bibnamefont  [1]{#1}%
\providecommand \bibfnamefont [1]{#1}%
\providecommand \citenamefont [1]{#1}%
\providecommand \href@noop [0]{\@secondoftwo}%
\providecommand \href [0]{\begingroup \@sanitize@url \@href}%
\providecommand \@href[1]{\@@startlink{#1}\@@href}%
\providecommand \@@href[1]{\endgroup#1\@@endlink}%
\providecommand \@sanitize@url [0]{\catcode `\\12\catcode `\$12\catcode `\&12\catcode `\#12\catcode `\^12\catcode `\_12\catcode `\%12\relax}%
\providecommand \@@startlink[1]{}%
\providecommand \@@endlink[0]{}%
\providecommand \url  [0]{\begingroup\@sanitize@url \@url }%
\providecommand \@url [1]{\endgroup\@href {#1}{\urlprefix }}%
\providecommand \urlprefix  [0]{URL }%
\providecommand \Eprint [0]{\href }%
\providecommand \doibase [0]{https://doi.org/}%
\providecommand \selectlanguage [0]{\@gobble}%
\providecommand \bibinfo  [0]{\@secondoftwo}%
\providecommand \bibfield  [0]{\@secondoftwo}%
\providecommand \translation [1]{[#1]}%
\providecommand \BibitemOpen [0]{}%
\providecommand \bibitemStop [0]{}%
\providecommand \bibitemNoStop [0]{.\EOS\space}%
\providecommand \EOS [0]{\spacefactor3000\relax}%
\providecommand \BibitemShut  [1]{\csname bibitem#1\endcsname}%
\let\auto@bib@innerbib\@empty
\bibitem [{\citenamefont {Or{\'u}s}(2019)}]{Orus2019}%
  \BibitemOpen
  \bibfield  {author} {\bibinfo {author} {\bibfnamefont {R.}~\bibnamefont {Or{\'u}s}},\ }\bibfield  {title} {\bibinfo {title} {Tensor networks for complex quantum systems},\ }\href {https://doi.org/10.1038/s42254-019-0086-7} {\bibfield  {journal} {\bibinfo  {journal} {Nature Reviews Physics}\ }\textbf {\bibinfo {volume} {1}},\ \bibinfo {pages} {538} (\bibinfo {year} {2019})}\BibitemShut {NoStop}%
\bibitem [{\citenamefont {Weimer}\ \emph {et~al.}(2021)\citenamefont {Weimer}, \citenamefont {Kshetrimayum},\ and\ \citenamefont {Or\'us}}]{Weimer2021}%
  \BibitemOpen
  \bibfield  {author} {\bibinfo {author} {\bibfnamefont {H.}~\bibnamefont {Weimer}}, \bibinfo {author} {\bibfnamefont {A.}~\bibnamefont {Kshetrimayum}},\ and\ \bibinfo {author} {\bibfnamefont {R.}~\bibnamefont {Or\'us}},\ }\bibfield  {title} {\bibinfo {title} {Simulation methods for open quantum many-body systems},\ }\href {https://doi.org/10.1103/RevModPhys.93.015008} {\bibfield  {journal} {\bibinfo  {journal} {Rev. Mod. Phys.}\ }\textbf {\bibinfo {volume} {93}},\ \bibinfo {pages} {015008} (\bibinfo {year} {2021})}\BibitemShut {NoStop}%
\bibitem [{\citenamefont {Cirac}\ \emph {et~al.}(2021)\citenamefont {Cirac}, \citenamefont {P\'erez-Garc\'{\i}a}, \citenamefont {Schuch},\ and\ \citenamefont {Verstraete}}]{Cirac2021}%
  \BibitemOpen
  \bibfield  {author} {\bibinfo {author} {\bibfnamefont {J.~I.}\ \bibnamefont {Cirac}}, \bibinfo {author} {\bibfnamefont {D.}~\bibnamefont {P\'erez-Garc\'{\i}a}}, \bibinfo {author} {\bibfnamefont {N.}~\bibnamefont {Schuch}},\ and\ \bibinfo {author} {\bibfnamefont {F.}~\bibnamefont {Verstraete}},\ }\bibfield  {title} {\bibinfo {title} {Matrix product states and projected entangled pair states: Concepts, symmetries, theorems},\ }\href {https://doi.org/10.1103/RevModPhys.93.045003} {\bibfield  {journal} {\bibinfo  {journal} {Rev. Mod. Phys.}\ }\textbf {\bibinfo {volume} {93}},\ \bibinfo {pages} {045003} (\bibinfo {year} {2021})}\BibitemShut {NoStop}%
\bibitem [{\citenamefont {Tindall}\ \emph {et~al.}(2024)\citenamefont {Tindall}, \citenamefont {Fishman}, \citenamefont {Stoudenmire},\ and\ \citenamefont {Sels}}]{Tindall23b}%
  \BibitemOpen
  \bibfield  {author} {\bibinfo {author} {\bibfnamefont {J.}~\bibnamefont {Tindall}}, \bibinfo {author} {\bibfnamefont {M.}~\bibnamefont {Fishman}}, \bibinfo {author} {\bibfnamefont {E.~M.}\ \bibnamefont {Stoudenmire}},\ and\ \bibinfo {author} {\bibfnamefont {D.}~\bibnamefont {Sels}},\ }\bibfield  {title} {\bibinfo {title} {Efficient tensor network simulation of ibm's eagle kicked ising experiment},\ }\href {https://doi.org/10.1103/PRXQuantum.5.010308} {\bibfield  {journal} {\bibinfo  {journal} {PRX Quantum}\ }\textbf {\bibinfo {volume} {5}},\ \bibinfo {pages} {010308} (\bibinfo {year} {2024})}\BibitemShut {NoStop}%
\bibitem [{\citenamefont {Patra}\ \emph {et~al.}(2024)\citenamefont {Patra}, \citenamefont {Jahromi}, \citenamefont {Singh},\ and\ \citenamefont {Or\'us}}]{Patra2024}%
  \BibitemOpen
  \bibfield  {author} {\bibinfo {author} {\bibfnamefont {S.}~\bibnamefont {Patra}}, \bibinfo {author} {\bibfnamefont {S.~S.}\ \bibnamefont {Jahromi}}, \bibinfo {author} {\bibfnamefont {S.}~\bibnamefont {Singh}},\ and\ \bibinfo {author} {\bibfnamefont {R.}~\bibnamefont {Or\'us}},\ }\bibfield  {title} {\bibinfo {title} {Efficient tensor network simulation of ibm's largest quantum processors},\ }\href {https://doi.org/10.1103/PhysRevResearch.6.013326} {\bibfield  {journal} {\bibinfo  {journal} {Phys. Rev. Res.}\ }\textbf {\bibinfo {volume} {6}},\ \bibinfo {pages} {013326} (\bibinfo {year} {2024})}\BibitemShut {NoStop}%
\bibitem [{\citenamefont {Sacha}\ and\ \citenamefont {Zakrzewski}(2017)}]{Sacha2018}%
  \BibitemOpen
  \bibfield  {author} {\bibinfo {author} {\bibfnamefont {K.}~\bibnamefont {Sacha}}\ and\ \bibinfo {author} {\bibfnamefont {J.}~\bibnamefont {Zakrzewski}},\ }\bibfield  {title} {\bibinfo {title} {Time crystals: a review},\ }\href {https://doi.org/10.1088/1361-6633/aa8b38} {\bibfield  {journal} {\bibinfo  {journal} {Reports on Progress in Physics}\ }\textbf {\bibinfo {volume} {81}},\ \bibinfo {pages} {016401} (\bibinfo {year} {2017})}\BibitemShut {NoStop}%
\bibitem [{\citenamefont {Khemani}\ \emph {et~al.}(2019)\citenamefont {Khemani}, \citenamefont {Moessner},\ and\ \citenamefont {Sondhi}}]{Khemani2019}%
  \BibitemOpen
  \bibfield  {author} {\bibinfo {author} {\bibfnamefont {V.}~\bibnamefont {Khemani}}, \bibinfo {author} {\bibfnamefont {R.}~\bibnamefont {Moessner}},\ and\ \bibinfo {author} {\bibfnamefont {S.~L.}\ \bibnamefont {Sondhi}},\ }\href {https://doi.org/https://doi.org/10.48550/arXiv.1910.10745} {\bibinfo {title} {A brief history of time crystals}} (\bibinfo {year} {2019}),\ \Eprint {https://arxiv.org/abs/1910.10745} {arXiv:1910.10745 [cond-mat.str-el]} \BibitemShut {NoStop}%
\bibitem [{\citenamefont {Else}\ \emph {et~al.}(2020)\citenamefont {Else}, \citenamefont {Monroe}, \citenamefont {Nayak},\ and\ \citenamefont {Yao}}]{Else2020}%
  \BibitemOpen
  \bibfield  {author} {\bibinfo {author} {\bibfnamefont {D.~V.}\ \bibnamefont {Else}}, \bibinfo {author} {\bibfnamefont {C.}~\bibnamefont {Monroe}}, \bibinfo {author} {\bibfnamefont {C.}~\bibnamefont {Nayak}},\ and\ \bibinfo {author} {\bibfnamefont {N.~Y.}\ \bibnamefont {Yao}},\ }\bibfield  {title} {\bibinfo {title} {Discrete time crystals},\ }\href {https://doi.org/10.1146/annurev-conmatphys-031119-050658} {\bibfield  {journal} {\bibinfo  {journal} {Annual Review of Condensed Matter Physics}\ }\textbf {\bibinfo {volume} {11}},\ \bibinfo {pages} {467–499} (\bibinfo {year} {2020})}\BibitemShut {NoStop}%
\bibitem [{\citenamefont {Guo}\ and\ \citenamefont {Liang}(2020)}]{Guo2020}%
  \BibitemOpen
  \bibfield  {author} {\bibinfo {author} {\bibfnamefont {L.}~\bibnamefont {Guo}}\ and\ \bibinfo {author} {\bibfnamefont {P.}~\bibnamefont {Liang}},\ }\bibfield  {title} {\bibinfo {title} {Condensed matter physics in time crystals},\ }\href {https://doi.org/10.1088/1367-2630/ab9d54} {\bibfield  {journal} {\bibinfo  {journal} {New Journal of Physics}\ }\textbf {\bibinfo {volume} {22}},\ \bibinfo {pages} {075003} (\bibinfo {year} {2020})}\BibitemShut {NoStop}%
\bibitem [{\citenamefont {Sacha}(2020)}]{Sacha2020}%
  \BibitemOpen
  \bibfield  {author} {\bibinfo {author} {\bibfnamefont {K.}~\bibnamefont {Sacha}},\ }\href {https://doi.org/https://doi.org/10.1007/978-3-030-52523-1} {\emph {\bibinfo {title} {Time Crystals}}}\ (\bibinfo  {publisher} {Springer Cham},\ \bibinfo {year} {2020})\BibitemShut {NoStop}%
\bibitem [{\citenamefont {Zaletel}\ \emph {et~al.}(2023)\citenamefont {Zaletel}, \citenamefont {Lukin}, \citenamefont {Monroe}, \citenamefont {Nayak}, \citenamefont {Wilczek},\ and\ \citenamefont {Yao}}]{Zaletel2023}%
  \BibitemOpen
  \bibfield  {author} {\bibinfo {author} {\bibfnamefont {M.~P.}\ \bibnamefont {Zaletel}}, \bibinfo {author} {\bibfnamefont {M.}~\bibnamefont {Lukin}}, \bibinfo {author} {\bibfnamefont {C.}~\bibnamefont {Monroe}}, \bibinfo {author} {\bibfnamefont {C.}~\bibnamefont {Nayak}}, \bibinfo {author} {\bibfnamefont {F.}~\bibnamefont {Wilczek}},\ and\ \bibinfo {author} {\bibfnamefont {N.~Y.}\ \bibnamefont {Yao}},\ }\bibfield  {title} {\bibinfo {title} {Colloquium: Quantum and classical discrete time crystals},\ }\href {https://doi.org/10.1103/RevModPhys.95.031001} {\bibfield  {journal} {\bibinfo  {journal} {Rev. Mod. Phys.}\ }\textbf {\bibinfo {volume} {95}},\ \bibinfo {pages} {031001} (\bibinfo {year} {2023})}\BibitemShut {NoStop}%
\bibitem [{\citenamefont {Else}\ \emph {et~al.}(2016)\citenamefont {Else}, \citenamefont {Bauer},\ and\ \citenamefont {Nayak}}]{Else2016}%
  \BibitemOpen
  \bibfield  {author} {\bibinfo {author} {\bibfnamefont {D.~V.}\ \bibnamefont {Else}}, \bibinfo {author} {\bibfnamefont {B.}~\bibnamefont {Bauer}},\ and\ \bibinfo {author} {\bibfnamefont {C.}~\bibnamefont {Nayak}},\ }\bibfield  {title} {\bibinfo {title} {Floquet time crystals},\ }\href {https://doi.org/10.1103/PhysRevLett.117.090402} {\bibfield  {journal} {\bibinfo  {journal} {Phys. Rev. Lett.}\ }\textbf {\bibinfo {volume} {117}},\ \bibinfo {pages} {090402} (\bibinfo {year} {2016})}\BibitemShut {NoStop}%
\bibitem [{\citenamefont {Khemani}\ \emph {et~al.}(2016)\citenamefont {Khemani}, \citenamefont {Lazarides}, \citenamefont {Moessner},\ and\ \citenamefont {Sondhi}}]{Khemani2016}%
  \BibitemOpen
  \bibfield  {author} {\bibinfo {author} {\bibfnamefont {V.}~\bibnamefont {Khemani}}, \bibinfo {author} {\bibfnamefont {A.}~\bibnamefont {Lazarides}}, \bibinfo {author} {\bibfnamefont {R.}~\bibnamefont {Moessner}},\ and\ \bibinfo {author} {\bibfnamefont {S.~L.}\ \bibnamefont {Sondhi}},\ }\bibfield  {title} {\bibinfo {title} {Phase structure of driven quantum systems},\ }\href {https://doi.org/10.1103/PhysRevLett.116.250401} {\bibfield  {journal} {\bibinfo  {journal} {Phys. Rev. Lett.}\ }\textbf {\bibinfo {volume} {116}},\ \bibinfo {pages} {250401} (\bibinfo {year} {2016})}\BibitemShut {NoStop}%
\bibitem [{\citenamefont {von Keyserlingk}\ \emph {et~al.}(2016)\citenamefont {von Keyserlingk}, \citenamefont {Khemani},\ and\ \citenamefont {Sondhi}}]{vonKeyserlingk2016}%
  \BibitemOpen
  \bibfield  {author} {\bibinfo {author} {\bibfnamefont {C.~W.}\ \bibnamefont {von Keyserlingk}}, \bibinfo {author} {\bibfnamefont {V.}~\bibnamefont {Khemani}},\ and\ \bibinfo {author} {\bibfnamefont {S.~L.}\ \bibnamefont {Sondhi}},\ }\bibfield  {title} {\bibinfo {title} {Absolute stability and spatiotemporal long-range order in floquet systems},\ }\href {https://doi.org/10.1103/PhysRevB.94.085112} {\bibfield  {journal} {\bibinfo  {journal} {Phys. Rev. B}\ }\textbf {\bibinfo {volume} {94}},\ \bibinfo {pages} {085112} (\bibinfo {year} {2016})}\BibitemShut {NoStop}%
\bibitem [{\citenamefont {Yao}\ \emph {et~al.}(2017)\citenamefont {Yao}, \citenamefont {Potter}, \citenamefont {Potirniche},\ and\ \citenamefont {Vishwanath}}]{Yao2017}%
  \BibitemOpen
  \bibfield  {author} {\bibinfo {author} {\bibfnamefont {N.~Y.}\ \bibnamefont {Yao}}, \bibinfo {author} {\bibfnamefont {A.~C.}\ \bibnamefont {Potter}}, \bibinfo {author} {\bibfnamefont {I.-D.}\ \bibnamefont {Potirniche}},\ and\ \bibinfo {author} {\bibfnamefont {A.}~\bibnamefont {Vishwanath}},\ }\bibfield  {title} {\bibinfo {title} {Discrete time crystals: Rigidity, criticality, and realizations},\ }\href {https://doi.org/10.1103/PhysRevLett.118.030401} {\bibfield  {journal} {\bibinfo  {journal} {Phys. Rev. Lett.}\ }\textbf {\bibinfo {volume} {118}},\ \bibinfo {pages} {030401} (\bibinfo {year} {2017})}\BibitemShut {NoStop}%
\bibitem [{\citenamefont {Zhang}\ \emph {et~al.}(2017)\citenamefont {Zhang}, \citenamefont {Hess}, \citenamefont {Kyprianidis}, \citenamefont {Becker}, \citenamefont {Lee}, \citenamefont {Smith}, \citenamefont {Pagano}, \citenamefont {Potirniche}, \citenamefont {Potter}, \citenamefont {Vishwanath}, \citenamefont {Yao},\ and\ \citenamefont {Monroe}}]{Zhang2017}%
  \BibitemOpen
  \bibfield  {author} {\bibinfo {author} {\bibfnamefont {J.}~\bibnamefont {Zhang}}, \bibinfo {author} {\bibfnamefont {P.~W.}\ \bibnamefont {Hess}}, \bibinfo {author} {\bibfnamefont {A.}~\bibnamefont {Kyprianidis}}, \bibinfo {author} {\bibfnamefont {P.}~\bibnamefont {Becker}}, \bibinfo {author} {\bibfnamefont {A.}~\bibnamefont {Lee}}, \bibinfo {author} {\bibfnamefont {J.}~\bibnamefont {Smith}}, \bibinfo {author} {\bibfnamefont {G.}~\bibnamefont {Pagano}}, \bibinfo {author} {\bibfnamefont {I.-D.}\ \bibnamefont {Potirniche}}, \bibinfo {author} {\bibfnamefont {A.~C.}\ \bibnamefont {Potter}}, \bibinfo {author} {\bibfnamefont {A.}~\bibnamefont {Vishwanath}}, \bibinfo {author} {\bibfnamefont {N.~Y.}\ \bibnamefont {Yao}},\ and\ \bibinfo {author} {\bibfnamefont {C.}~\bibnamefont {Monroe}},\ }\bibfield  {title} {\bibinfo {title} {Observation of a discrete time crystal},\ }\href {https://doi.org/10.1038/nature21413} {\bibfield  {journal} {\bibinfo  {journal} {Nature}\ }\textbf {\bibinfo {volume} {543}},\ \bibinfo
  {pages} {217–220} (\bibinfo {year} {2017})}\BibitemShut {NoStop}%
\bibitem [{\citenamefont {Ippoliti}\ \emph {et~al.}(2021)\citenamefont {Ippoliti}, \citenamefont {Kechedzhi}, \citenamefont {Moessner}, \citenamefont {Sondhi},\ and\ \citenamefont {Khemani}}]{Ippoliti2021}%
  \BibitemOpen
  \bibfield  {author} {\bibinfo {author} {\bibfnamefont {M.}~\bibnamefont {Ippoliti}}, \bibinfo {author} {\bibfnamefont {K.}~\bibnamefont {Kechedzhi}}, \bibinfo {author} {\bibfnamefont {R.}~\bibnamefont {Moessner}}, \bibinfo {author} {\bibfnamefont {S.}~\bibnamefont {Sondhi}},\ and\ \bibinfo {author} {\bibfnamefont {V.}~\bibnamefont {Khemani}},\ }\bibfield  {title} {\bibinfo {title} {Many-body physics in the nisq era: Quantum programming a discrete time crystal},\ }\href {https://doi.org/10.1103/PRXQuantum.2.030346} {\bibfield  {journal} {\bibinfo  {journal} {PRX Quantum}\ }\textbf {\bibinfo {volume} {2}},\ \bibinfo {pages} {030346} (\bibinfo {year} {2021})}\BibitemShut {NoStop}%
\bibitem [{\citenamefont {Mi}\ \emph {et~al.}(2022)\citenamefont {Mi}, \citenamefont {Ippoliti}, \citenamefont {Quintana}, \citenamefont {Greene}, \citenamefont {Chen}, \citenamefont {Gross}, \citenamefont {Arute}, \citenamefont {Arya}, \citenamefont {Atalaya}, \citenamefont {Babbush}, \citenamefont {Bardin}, \citenamefont {Basso}, \citenamefont {Bengtsson}, \citenamefont {Bilmes}, \citenamefont {Bourassa}, \citenamefont {Brill}, \citenamefont {Broughton}, \citenamefont {Buckley}, \citenamefont {Buell}, \citenamefont {Burkett}, \citenamefont {Bushnell}, \citenamefont {Chiaro}, \citenamefont {Collins}, \citenamefont {Courtney}, \citenamefont {Debroy}, \citenamefont {Demura}, \citenamefont {Derk}, \citenamefont {Dunsworth}, \citenamefont {Eppens}, \citenamefont {Erickson}, \citenamefont {Farhi}, \citenamefont {Fowler}, \citenamefont {Foxen}, \citenamefont {Gidney}, \citenamefont {Giustina}, \citenamefont {Harrigan}, \citenamefont {Harrington}, \citenamefont {Hilton}, \citenamefont {Ho}, \citenamefont {Hong},
  \citenamefont {Huang}, \citenamefont {Huff}, \citenamefont {Huggins}, \citenamefont {Ioffe}, \citenamefont {Isakov}, \citenamefont {Iveland}, \citenamefont {Jeffrey}, \citenamefont {Jiang}, \citenamefont {Jones}, \citenamefont {Kafri}, \citenamefont {Khattar}, \citenamefont {Kim}, \citenamefont {Kitaev}, \citenamefont {Klimov}, \citenamefont {Korotkov}, \citenamefont {Kostritsa}, \citenamefont {Landhuis}, \citenamefont {Laptev}, \citenamefont {Lee}, \citenamefont {Lee}, \citenamefont {Locharla}, \citenamefont {Lucero}, \citenamefont {Martin}, \citenamefont {McClean}, \citenamefont {McCourt}, \citenamefont {McEwen}, \citenamefont {Miao}, \citenamefont {Mohseni}, \citenamefont {Montazeri}, \citenamefont {Mruczkiewicz}, \citenamefont {Naaman}, \citenamefont {Neeley}, \citenamefont {Neill}, \citenamefont {Newman}, \citenamefont {Niu}, \citenamefont {O’Brien}, \citenamefont {Opremcak}, \citenamefont {Ostby}, \citenamefont {Pato}, \citenamefont {Petukhov}, \citenamefont {Rubin}, \citenamefont {Sank},
  \citenamefont {Satzinger}, \citenamefont {Shvarts}, \citenamefont {Su}, \citenamefont {Strain}, \citenamefont {Szalay}, \citenamefont {Trevithick}, \citenamefont {Villalonga}, \citenamefont {White}, \citenamefont {Yao}, \citenamefont {Yeh}, \citenamefont {Yoo}, \citenamefont {Zalcman}, \citenamefont {Neven}, \citenamefont {Boixo}, \citenamefont {Smelyanskiy}, \citenamefont {Megrant}, \citenamefont {Kelly}, \citenamefont {Chen}, \citenamefont {Sondhi}, \citenamefont {Moessner}, \citenamefont {Kechedzhi}, \citenamefont {Khemani},\ and\ \citenamefont {Roushan}}]{Mi2022}%
  \BibitemOpen
  \bibfield  {author} {\bibinfo {author} {\bibfnamefont {X.}~\bibnamefont {Mi}}, \bibinfo {author} {\bibfnamefont {M.}~\bibnamefont {Ippoliti}}, \bibinfo {author} {\bibfnamefont {C.}~\bibnamefont {Quintana}}, \bibinfo {author} {\bibfnamefont {A.}~\bibnamefont {Greene}}, \bibinfo {author} {\bibfnamefont {Z.}~\bibnamefont {Chen}}, \bibinfo {author} {\bibfnamefont {J.}~\bibnamefont {Gross}}, \bibinfo {author} {\bibfnamefont {F.}~\bibnamefont {Arute}}, \bibinfo {author} {\bibfnamefont {K.}~\bibnamefont {Arya}}, \bibinfo {author} {\bibfnamefont {J.}~\bibnamefont {Atalaya}}, \bibinfo {author} {\bibfnamefont {R.}~\bibnamefont {Babbush}}, \bibinfo {author} {\bibfnamefont {J.~C.}\ \bibnamefont {Bardin}}, \bibinfo {author} {\bibfnamefont {J.}~\bibnamefont {Basso}}, \bibinfo {author} {\bibfnamefont {A.}~\bibnamefont {Bengtsson}}, \bibinfo {author} {\bibfnamefont {A.}~\bibnamefont {Bilmes}}, \bibinfo {author} {\bibfnamefont {A.}~\bibnamefont {Bourassa}}, \bibinfo {author} {\bibfnamefont {L.}~\bibnamefont {Brill}},
  \bibinfo {author} {\bibfnamefont {M.}~\bibnamefont {Broughton}}, \bibinfo {author} {\bibfnamefont {B.~B.}\ \bibnamefont {Buckley}}, \bibinfo {author} {\bibfnamefont {D.~A.}\ \bibnamefont {Buell}}, \bibinfo {author} {\bibfnamefont {B.}~\bibnamefont {Burkett}}, \bibinfo {author} {\bibfnamefont {N.}~\bibnamefont {Bushnell}}, \bibinfo {author} {\bibfnamefont {B.}~\bibnamefont {Chiaro}}, \bibinfo {author} {\bibfnamefont {R.}~\bibnamefont {Collins}}, \bibinfo {author} {\bibfnamefont {W.}~\bibnamefont {Courtney}}, \bibinfo {author} {\bibfnamefont {D.}~\bibnamefont {Debroy}}, \bibinfo {author} {\bibfnamefont {S.}~\bibnamefont {Demura}}, \bibinfo {author} {\bibfnamefont {A.~R.}\ \bibnamefont {Derk}}, \bibinfo {author} {\bibfnamefont {A.}~\bibnamefont {Dunsworth}}, \bibinfo {author} {\bibfnamefont {D.}~\bibnamefont {Eppens}}, \bibinfo {author} {\bibfnamefont {C.}~\bibnamefont {Erickson}}, \bibinfo {author} {\bibfnamefont {E.}~\bibnamefont {Farhi}}, \bibinfo {author} {\bibfnamefont {A.~G.}\ \bibnamefont {Fowler}},
  \bibinfo {author} {\bibfnamefont {B.}~\bibnamefont {Foxen}}, \bibinfo {author} {\bibfnamefont {C.}~\bibnamefont {Gidney}}, \bibinfo {author} {\bibfnamefont {M.}~\bibnamefont {Giustina}}, \bibinfo {author} {\bibfnamefont {M.~P.}\ \bibnamefont {Harrigan}}, \bibinfo {author} {\bibfnamefont {S.~D.}\ \bibnamefont {Harrington}}, \bibinfo {author} {\bibfnamefont {J.}~\bibnamefont {Hilton}}, \bibinfo {author} {\bibfnamefont {A.}~\bibnamefont {Ho}}, \bibinfo {author} {\bibfnamefont {S.}~\bibnamefont {Hong}}, \bibinfo {author} {\bibfnamefont {T.}~\bibnamefont {Huang}}, \bibinfo {author} {\bibfnamefont {A.}~\bibnamefont {Huff}}, \bibinfo {author} {\bibfnamefont {W.~J.}\ \bibnamefont {Huggins}}, \bibinfo {author} {\bibfnamefont {L.~B.}\ \bibnamefont {Ioffe}}, \bibinfo {author} {\bibfnamefont {S.~V.}\ \bibnamefont {Isakov}}, \bibinfo {author} {\bibfnamefont {J.}~\bibnamefont {Iveland}}, \bibinfo {author} {\bibfnamefont {E.}~\bibnamefont {Jeffrey}}, \bibinfo {author} {\bibfnamefont {Z.}~\bibnamefont {Jiang}}, \bibinfo
  {author} {\bibfnamefont {C.}~\bibnamefont {Jones}}, \bibinfo {author} {\bibfnamefont {D.}~\bibnamefont {Kafri}}, \bibinfo {author} {\bibfnamefont {T.}~\bibnamefont {Khattar}}, \bibinfo {author} {\bibfnamefont {S.}~\bibnamefont {Kim}}, \bibinfo {author} {\bibfnamefont {A.}~\bibnamefont {Kitaev}}, \bibinfo {author} {\bibfnamefont {P.~V.}\ \bibnamefont {Klimov}}, \bibinfo {author} {\bibfnamefont {A.~N.}\ \bibnamefont {Korotkov}}, \bibinfo {author} {\bibfnamefont {F.}~\bibnamefont {Kostritsa}}, \bibinfo {author} {\bibfnamefont {D.}~\bibnamefont {Landhuis}}, \bibinfo {author} {\bibfnamefont {P.}~\bibnamefont {Laptev}}, \bibinfo {author} {\bibfnamefont {J.}~\bibnamefont {Lee}}, \bibinfo {author} {\bibfnamefont {K.}~\bibnamefont {Lee}}, \bibinfo {author} {\bibfnamefont {A.}~\bibnamefont {Locharla}}, \bibinfo {author} {\bibfnamefont {E.}~\bibnamefont {Lucero}}, \bibinfo {author} {\bibfnamefont {O.}~\bibnamefont {Martin}}, \bibinfo {author} {\bibfnamefont {J.~R.}\ \bibnamefont {McClean}}, \bibinfo {author}
  {\bibfnamefont {T.}~\bibnamefont {McCourt}}, \bibinfo {author} {\bibfnamefont {M.}~\bibnamefont {McEwen}}, \bibinfo {author} {\bibfnamefont {K.~C.}\ \bibnamefont {Miao}}, \bibinfo {author} {\bibfnamefont {M.}~\bibnamefont {Mohseni}}, \bibinfo {author} {\bibfnamefont {S.}~\bibnamefont {Montazeri}}, \bibinfo {author} {\bibfnamefont {W.}~\bibnamefont {Mruczkiewicz}}, \bibinfo {author} {\bibfnamefont {O.}~\bibnamefont {Naaman}}, \bibinfo {author} {\bibfnamefont {M.}~\bibnamefont {Neeley}}, \bibinfo {author} {\bibfnamefont {C.}~\bibnamefont {Neill}}, \bibinfo {author} {\bibfnamefont {M.}~\bibnamefont {Newman}}, \bibinfo {author} {\bibfnamefont {M.~Y.}\ \bibnamefont {Niu}}, \bibinfo {author} {\bibfnamefont {T.~E.}\ \bibnamefont {O’Brien}}, \bibinfo {author} {\bibfnamefont {A.}~\bibnamefont {Opremcak}}, \bibinfo {author} {\bibfnamefont {E.}~\bibnamefont {Ostby}}, \bibinfo {author} {\bibfnamefont {B.}~\bibnamefont {Pato}}, \bibinfo {author} {\bibfnamefont {A.}~\bibnamefont {Petukhov}}, \bibinfo {author}
  {\bibfnamefont {N.~C.}\ \bibnamefont {Rubin}}, \bibinfo {author} {\bibfnamefont {D.}~\bibnamefont {Sank}}, \bibinfo {author} {\bibfnamefont {K.~J.}\ \bibnamefont {Satzinger}}, \bibinfo {author} {\bibfnamefont {V.}~\bibnamefont {Shvarts}}, \bibinfo {author} {\bibfnamefont {Y.}~\bibnamefont {Su}}, \bibinfo {author} {\bibfnamefont {D.}~\bibnamefont {Strain}}, \bibinfo {author} {\bibfnamefont {M.}~\bibnamefont {Szalay}}, \bibinfo {author} {\bibfnamefont {M.~D.}\ \bibnamefont {Trevithick}}, \bibinfo {author} {\bibfnamefont {B.}~\bibnamefont {Villalonga}}, \bibinfo {author} {\bibfnamefont {T.}~\bibnamefont {White}}, \bibinfo {author} {\bibfnamefont {Z.~J.}\ \bibnamefont {Yao}}, \bibinfo {author} {\bibfnamefont {P.}~\bibnamefont {Yeh}}, \bibinfo {author} {\bibfnamefont {J.}~\bibnamefont {Yoo}}, \bibinfo {author} {\bibfnamefont {A.}~\bibnamefont {Zalcman}}, \bibinfo {author} {\bibfnamefont {H.}~\bibnamefont {Neven}}, \bibinfo {author} {\bibfnamefont {S.}~\bibnamefont {Boixo}}, \bibinfo {author} {\bibfnamefont
  {V.}~\bibnamefont {Smelyanskiy}}, \bibinfo {author} {\bibfnamefont {A.}~\bibnamefont {Megrant}}, \bibinfo {author} {\bibfnamefont {J.}~\bibnamefont {Kelly}}, \bibinfo {author} {\bibfnamefont {Y.}~\bibnamefont {Chen}}, \bibinfo {author} {\bibfnamefont {S.~L.}\ \bibnamefont {Sondhi}}, \bibinfo {author} {\bibfnamefont {R.}~\bibnamefont {Moessner}}, \bibinfo {author} {\bibfnamefont {K.}~\bibnamefont {Kechedzhi}}, \bibinfo {author} {\bibfnamefont {V.}~\bibnamefont {Khemani}},\ and\ \bibinfo {author} {\bibfnamefont {P.}~\bibnamefont {Roushan}},\ }\bibfield  {title} {\bibinfo {title} {Time-crystalline eigenstate order on a quantum processor},\ }\href {https://doi.org/10.1038/s41586-021-04257-w} {\bibfield  {journal} {\bibinfo  {journal} {Nature}\ }\textbf {\bibinfo {volume} {601}},\ \bibinfo {pages} {531–536} (\bibinfo {year} {2022})}\BibitemShut {NoStop}%
\bibitem [{\citenamefont {Frey}\ and\ \citenamefont {Rachel}(2022)}]{Frey2022}%
  \BibitemOpen
  \bibfield  {author} {\bibinfo {author} {\bibfnamefont {P.}~\bibnamefont {Frey}}\ and\ \bibinfo {author} {\bibfnamefont {S.}~\bibnamefont {Rachel}},\ }\bibfield  {title} {\bibinfo {title} {Realization of a discrete time crystal on 57 qubits of a quantum computer},\ }\href {https://doi.org/10.1126/sciadv.abm7652} {\bibfield  {journal} {\bibinfo  {journal} {Science Advances}\ }\textbf {\bibinfo {volume} {8}},\ \bibinfo {pages} {eabm7652} (\bibinfo {year} {2022})},\ \Eprint {https://arxiv.org/abs/https://www.science.org/doi/pdf/10.1126/sciadv.abm7652} {https://www.science.org/doi/pdf/10.1126/sciadv.abm7652} \BibitemShut {NoStop}%
\bibitem [{\citenamefont {Zhang}\ \emph {et~al.}(2022)\citenamefont {Zhang}, \citenamefont {Jiang}, \citenamefont {Deng}, \citenamefont {Wang}, \citenamefont {Chen}, \citenamefont {Zhang}, \citenamefont {Ren}, \citenamefont {Dong}, \citenamefont {Xu}, \citenamefont {Gao}, \citenamefont {Jin}, \citenamefont {Zhu}, \citenamefont {Guo}, \citenamefont {Li}, \citenamefont {Song}, \citenamefont {Gorshkov}, \citenamefont {Iadecola}, \citenamefont {Liu}, \citenamefont {Gong}, \citenamefont {Wang}, \citenamefont {Deng},\ and\ \citenamefont {Wang}}]{Zhang2022}%
  \BibitemOpen
  \bibfield  {author} {\bibinfo {author} {\bibfnamefont {X.}~\bibnamefont {Zhang}}, \bibinfo {author} {\bibfnamefont {W.}~\bibnamefont {Jiang}}, \bibinfo {author} {\bibfnamefont {J.}~\bibnamefont {Deng}}, \bibinfo {author} {\bibfnamefont {K.}~\bibnamefont {Wang}}, \bibinfo {author} {\bibfnamefont {J.}~\bibnamefont {Chen}}, \bibinfo {author} {\bibfnamefont {P.}~\bibnamefont {Zhang}}, \bibinfo {author} {\bibfnamefont {W.}~\bibnamefont {Ren}}, \bibinfo {author} {\bibfnamefont {H.}~\bibnamefont {Dong}}, \bibinfo {author} {\bibfnamefont {S.}~\bibnamefont {Xu}}, \bibinfo {author} {\bibfnamefont {Y.}~\bibnamefont {Gao}}, \bibinfo {author} {\bibfnamefont {F.}~\bibnamefont {Jin}}, \bibinfo {author} {\bibfnamefont {X.}~\bibnamefont {Zhu}}, \bibinfo {author} {\bibfnamefont {Q.}~\bibnamefont {Guo}}, \bibinfo {author} {\bibfnamefont {H.}~\bibnamefont {Li}}, \bibinfo {author} {\bibfnamefont {C.}~\bibnamefont {Song}}, \bibinfo {author} {\bibfnamefont {A.~V.}\ \bibnamefont {Gorshkov}}, \bibinfo {author} {\bibfnamefont
  {T.}~\bibnamefont {Iadecola}}, \bibinfo {author} {\bibfnamefont {F.}~\bibnamefont {Liu}}, \bibinfo {author} {\bibfnamefont {Z.-X.}\ \bibnamefont {Gong}}, \bibinfo {author} {\bibfnamefont {Z.}~\bibnamefont {Wang}}, \bibinfo {author} {\bibfnamefont {D.-L.}\ \bibnamefont {Deng}},\ and\ \bibinfo {author} {\bibfnamefont {H.}~\bibnamefont {Wang}},\ }\bibfield  {title} {\bibinfo {title} {Digital quantum simulation of floquet symmetry-protected topological phases},\ }\href {https://doi.org/https://doi.org/10.5281/zenodo.6510867} {\bibfield  {journal} {\bibinfo  {journal} {Nature}\ }\textbf {\bibinfo {volume} {607}},\ \bibinfo {pages} {468} (\bibinfo {year} {2022})}\BibitemShut {NoStop}%
\bibitem [{\citenamefont {Else}\ \emph {et~al.}(2017)\citenamefont {Else}, \citenamefont {Bauer},\ and\ \citenamefont {Nayak}}]{Else2017}%
  \BibitemOpen
  \bibfield  {author} {\bibinfo {author} {\bibfnamefont {D.~V.}\ \bibnamefont {Else}}, \bibinfo {author} {\bibfnamefont {B.}~\bibnamefont {Bauer}},\ and\ \bibinfo {author} {\bibfnamefont {C.}~\bibnamefont {Nayak}},\ }\bibfield  {title} {\bibinfo {title} {Prethermal phases of matter protected by time-translation symmetry},\ }\href {https://doi.org/10.1103/PhysRevX.7.011026} {\bibfield  {journal} {\bibinfo  {journal} {Phys. Rev. X}\ }\textbf {\bibinfo {volume} {7}},\ \bibinfo {pages} {011026} (\bibinfo {year} {2017})}\BibitemShut {NoStop}%
\bibitem [{\citenamefont {Machado}\ \emph {et~al.}(2020)\citenamefont {Machado}, \citenamefont {Else}, \citenamefont {Kahanamoku-Meyer}, \citenamefont {Nayak},\ and\ \citenamefont {Yao}}]{Machado2020}%
  \BibitemOpen
  \bibfield  {author} {\bibinfo {author} {\bibfnamefont {F.}~\bibnamefont {Machado}}, \bibinfo {author} {\bibfnamefont {D.~V.}\ \bibnamefont {Else}}, \bibinfo {author} {\bibfnamefont {G.~D.}\ \bibnamefont {Kahanamoku-Meyer}}, \bibinfo {author} {\bibfnamefont {C.}~\bibnamefont {Nayak}},\ and\ \bibinfo {author} {\bibfnamefont {N.~Y.}\ \bibnamefont {Yao}},\ }\bibfield  {title} {\bibinfo {title} {Long-range prethermal phases of nonequilibrium matter},\ }\href {https://doi.org/10.1103/PhysRevX.10.011043} {\bibfield  {journal} {\bibinfo  {journal} {Phys. Rev. X}\ }\textbf {\bibinfo {volume} {10}},\ \bibinfo {pages} {011043} (\bibinfo {year} {2020})}\BibitemShut {NoStop}%
\bibitem [{\citenamefont {Kyprianidis}\ \emph {et~al.}(2021)\citenamefont {Kyprianidis}, \citenamefont {Machado}, \citenamefont {Morong}, \citenamefont {Becker}, \citenamefont {Collins}, \citenamefont {Else}, \citenamefont {Feng}, \citenamefont {Hess}, \citenamefont {Nayak}, \citenamefont {Pagano}, \citenamefont {Yao},\ and\ \citenamefont {Monroe}}]{Kyprianidis2021}%
  \BibitemOpen
  \bibfield  {author} {\bibinfo {author} {\bibfnamefont {A.}~\bibnamefont {Kyprianidis}}, \bibinfo {author} {\bibfnamefont {F.}~\bibnamefont {Machado}}, \bibinfo {author} {\bibfnamefont {W.}~\bibnamefont {Morong}}, \bibinfo {author} {\bibfnamefont {P.}~\bibnamefont {Becker}}, \bibinfo {author} {\bibfnamefont {K.~S.}\ \bibnamefont {Collins}}, \bibinfo {author} {\bibfnamefont {D.~V.}\ \bibnamefont {Else}}, \bibinfo {author} {\bibfnamefont {L.}~\bibnamefont {Feng}}, \bibinfo {author} {\bibfnamefont {P.~W.}\ \bibnamefont {Hess}}, \bibinfo {author} {\bibfnamefont {C.}~\bibnamefont {Nayak}}, \bibinfo {author} {\bibfnamefont {G.}~\bibnamefont {Pagano}}, \bibinfo {author} {\bibfnamefont {N.~Y.}\ \bibnamefont {Yao}},\ and\ \bibinfo {author} {\bibfnamefont {C.}~\bibnamefont {Monroe}},\ }\bibfield  {title} {\bibinfo {title} {Observation of a prethermal discrete time crystal},\ }\href {https://doi.org/10.1126/science.abg8102} {\bibfield  {journal} {\bibinfo  {journal} {Science}\ }\textbf {\bibinfo {volume} {372}},\
  \bibinfo {pages} {1192–1196} (\bibinfo {year} {2021})}\BibitemShut {NoStop}%
\bibitem [{\citenamefont {Beatrez}\ \emph {et~al.}(2023)\citenamefont {Beatrez}, \citenamefont {Fleckenstein}, \citenamefont {Pillai}, \citenamefont {de~Leon~Sanchez}, \citenamefont {Akkiraju}, \citenamefont {Diaz~Alcala}, \citenamefont {Conti}, \citenamefont {Reshetikhin}, \citenamefont {Druga}, \citenamefont {Bukov},\ and\ \citenamefont {Ajoy}}]{Beatrez2023}%
  \BibitemOpen
  \bibfield  {author} {\bibinfo {author} {\bibfnamefont {W.}~\bibnamefont {Beatrez}}, \bibinfo {author} {\bibfnamefont {C.}~\bibnamefont {Fleckenstein}}, \bibinfo {author} {\bibfnamefont {A.}~\bibnamefont {Pillai}}, \bibinfo {author} {\bibfnamefont {E.}~\bibnamefont {de~Leon~Sanchez}}, \bibinfo {author} {\bibfnamefont {A.}~\bibnamefont {Akkiraju}}, \bibinfo {author} {\bibfnamefont {J.}~\bibnamefont {Diaz~Alcala}}, \bibinfo {author} {\bibfnamefont {S.}~\bibnamefont {Conti}}, \bibinfo {author} {\bibfnamefont {P.}~\bibnamefont {Reshetikhin}}, \bibinfo {author} {\bibfnamefont {E.}~\bibnamefont {Druga}}, \bibinfo {author} {\bibfnamefont {M.}~\bibnamefont {Bukov}},\ and\ \bibinfo {author} {\bibfnamefont {A.}~\bibnamefont {Ajoy}},\ }\bibfield  {title} {\bibinfo {title} {Critical prethermal discrete time crystal created by two-frequency driving},\ }\href {https://doi.org/10.1038/s41567-022-01891-7} {\bibfield  {journal} {\bibinfo  {journal} {Nature Physics}\ }\textbf {\bibinfo {volume} {19}},\ \bibinfo {pages} {407}
  (\bibinfo {year} {2023})}\BibitemShut {NoStop}%
\bibitem [{\citenamefont {Jin}\ \emph {et~al.}(2025)\citenamefont {Jin}, \citenamefont {Jiang}, \citenamefont {Zhu}, \citenamefont {Bao}, \citenamefont {Shen}, \citenamefont {Wang}, \citenamefont {Zhu}, \citenamefont {Xu}, \citenamefont {Song}, \citenamefont {Chen}, \citenamefont {Tan}, \citenamefont {Wu}, \citenamefont {Zhang}, \citenamefont {Gao}, \citenamefont {Wang}, \citenamefont {Zou}, \citenamefont {Zhang}, \citenamefont {Li}, \citenamefont {Zhong}, \citenamefont {Cui}, \citenamefont {Han}, \citenamefont {He}, \citenamefont {Wang}, \citenamefont {Yang}, \citenamefont {Wang}, \citenamefont {Shen}, \citenamefont {Liu}, \citenamefont {Deng}, \citenamefont {Dong}, \citenamefont {Zhang}, \citenamefont {Li}, \citenamefont {Yuan}, \citenamefont {Lu}, \citenamefont {Sun}, \citenamefont {Li}, \citenamefont {Zhang}, \citenamefont {Song}, \citenamefont {Wang}, \citenamefont {Guo}, \citenamefont {Machado}, \citenamefont {Kemp}, \citenamefont {Iadecola}, \citenamefont {Yao}, \citenamefont {Wang},\ and\
  \citenamefont {Deng}}]{Jin2025}%
  \BibitemOpen
  \bibfield  {author} {\bibinfo {author} {\bibfnamefont {F.}~\bibnamefont {Jin}}, \bibinfo {author} {\bibfnamefont {S.}~\bibnamefont {Jiang}}, \bibinfo {author} {\bibfnamefont {X.}~\bibnamefont {Zhu}}, \bibinfo {author} {\bibfnamefont {Z.}~\bibnamefont {Bao}}, \bibinfo {author} {\bibfnamefont {F.}~\bibnamefont {Shen}}, \bibinfo {author} {\bibfnamefont {K.}~\bibnamefont {Wang}}, \bibinfo {author} {\bibfnamefont {Z.}~\bibnamefont {Zhu}}, \bibinfo {author} {\bibfnamefont {S.}~\bibnamefont {Xu}}, \bibinfo {author} {\bibfnamefont {Z.}~\bibnamefont {Song}}, \bibinfo {author} {\bibfnamefont {J.}~\bibnamefont {Chen}}, \bibinfo {author} {\bibfnamefont {Z.}~\bibnamefont {Tan}}, \bibinfo {author} {\bibfnamefont {Y.}~\bibnamefont {Wu}}, \bibinfo {author} {\bibfnamefont {C.}~\bibnamefont {Zhang}}, \bibinfo {author} {\bibfnamefont {Y.}~\bibnamefont {Gao}}, \bibinfo {author} {\bibfnamefont {N.}~\bibnamefont {Wang}}, \bibinfo {author} {\bibfnamefont {Y.}~\bibnamefont {Zou}}, \bibinfo {author} {\bibfnamefont {A.}~\bibnamefont
  {Zhang}}, \bibinfo {author} {\bibfnamefont {T.}~\bibnamefont {Li}}, \bibinfo {author} {\bibfnamefont {J.}~\bibnamefont {Zhong}}, \bibinfo {author} {\bibfnamefont {Z.}~\bibnamefont {Cui}}, \bibinfo {author} {\bibfnamefont {Y.}~\bibnamefont {Han}}, \bibinfo {author} {\bibfnamefont {Y.}~\bibnamefont {He}}, \bibinfo {author} {\bibfnamefont {H.}~\bibnamefont {Wang}}, \bibinfo {author} {\bibfnamefont {J.-N.}\ \bibnamefont {Yang}}, \bibinfo {author} {\bibfnamefont {Y.}~\bibnamefont {Wang}}, \bibinfo {author} {\bibfnamefont {J.}~\bibnamefont {Shen}}, \bibinfo {author} {\bibfnamefont {G.}~\bibnamefont {Liu}}, \bibinfo {author} {\bibfnamefont {J.}~\bibnamefont {Deng}}, \bibinfo {author} {\bibfnamefont {H.}~\bibnamefont {Dong}}, \bibinfo {author} {\bibfnamefont {P.}~\bibnamefont {Zhang}}, \bibinfo {author} {\bibfnamefont {W.}~\bibnamefont {Li}}, \bibinfo {author} {\bibfnamefont {D.}~\bibnamefont {Yuan}}, \bibinfo {author} {\bibfnamefont {Z.}~\bibnamefont {Lu}}, \bibinfo {author} {\bibfnamefont {Z.-Z.}\ \bibnamefont
  {Sun}}, \bibinfo {author} {\bibfnamefont {H.}~\bibnamefont {Li}}, \bibinfo {author} {\bibfnamefont {J.}~\bibnamefont {Zhang}}, \bibinfo {author} {\bibfnamefont {C.}~\bibnamefont {Song}}, \bibinfo {author} {\bibfnamefont {Z.}~\bibnamefont {Wang}}, \bibinfo {author} {\bibfnamefont {Q.}~\bibnamefont {Guo}}, \bibinfo {author} {\bibfnamefont {F.}~\bibnamefont {Machado}}, \bibinfo {author} {\bibfnamefont {J.}~\bibnamefont {Kemp}}, \bibinfo {author} {\bibfnamefont {T.}~\bibnamefont {Iadecola}}, \bibinfo {author} {\bibfnamefont {N.~Y.}\ \bibnamefont {Yao}}, \bibinfo {author} {\bibfnamefont {H.}~\bibnamefont {Wang}},\ and\ \bibinfo {author} {\bibfnamefont {D.-L.}\ \bibnamefont {Deng}},\ }\bibfield  {title} {\bibinfo {title} {Topological prethermal strong zero modes on superconducting processors},\ }\href {https://doi.org/10.1038/s41586-025-09476-z} {\bibfield  {journal} {\bibinfo  {journal} {Nature}\ }\textbf {\bibinfo {volume} {645}},\ \bibinfo {pages} {626} (\bibinfo {year} {2025})}\BibitemShut {NoStop}%
\bibitem [{\citenamefont {Jiang}\ \emph {et~al.}(2025)\citenamefont {Jiang}, \citenamefont {Yuan}, \citenamefont {Jiang}, \citenamefont {Deng},\ and\ \citenamefont {Machado}}]{Jiang2025}%
  \BibitemOpen
  \bibfield  {author} {\bibinfo {author} {\bibfnamefont {S.}~\bibnamefont {Jiang}}, \bibinfo {author} {\bibfnamefont {D.}~\bibnamefont {Yuan}}, \bibinfo {author} {\bibfnamefont {W.}~\bibnamefont {Jiang}}, \bibinfo {author} {\bibfnamefont {D.-L.}\ \bibnamefont {Deng}},\ and\ \bibinfo {author} {\bibfnamefont {F.}~\bibnamefont {Machado}},\ }\bibfield  {title} {\bibinfo {title} {Prethermal time-crystalline corner modes},\ }\href {https://doi.org/10.1103/np9w-jsf9} {\bibfield  {journal} {\bibinfo  {journal} {Phys. Rev. Lett.}\ }\textbf {\bibinfo {volume} {135}},\ \bibinfo {pages} {110401} (\bibinfo {year} {2025})}\BibitemShut {NoStop}%
\bibitem [{\citenamefont {Huang}\ \emph {et~al.}(2018)\citenamefont {Huang}, \citenamefont {Wu},\ and\ \citenamefont {Liu}}]{Huang2018}%
  \BibitemOpen
  \bibfield  {author} {\bibinfo {author} {\bibfnamefont {B.}~\bibnamefont {Huang}}, \bibinfo {author} {\bibfnamefont {Y.-H.}\ \bibnamefont {Wu}},\ and\ \bibinfo {author} {\bibfnamefont {W.~V.}\ \bibnamefont {Liu}},\ }\bibfield  {title} {\bibinfo {title} {Clean floquet time crystals: Models and realizations in cold atoms},\ }\href {https://doi.org/10.1103/PhysRevLett.120.110603} {\bibfield  {journal} {\bibinfo  {journal} {Phys. Rev. Lett.}\ }\textbf {\bibinfo {volume} {120}},\ \bibinfo {pages} {110603} (\bibinfo {year} {2018})}\BibitemShut {NoStop}%
\bibitem [{\citenamefont {Pizzi}\ \emph {et~al.}(2021)\citenamefont {Pizzi}, \citenamefont {Knolle},\ and\ \citenamefont {Nunnenkamp}}]{Pizzi2021}%
  \BibitemOpen
  \bibfield  {author} {\bibinfo {author} {\bibfnamefont {A.}~\bibnamefont {Pizzi}}, \bibinfo {author} {\bibfnamefont {J.}~\bibnamefont {Knolle}},\ and\ \bibinfo {author} {\bibfnamefont {A.}~\bibnamefont {Nunnenkamp}},\ }\bibfield  {title} {\bibinfo {title} {Higher-order and fractional discrete time crystals in clean long-range interacting systems},\ }\bibfield  {journal} {\bibinfo  {journal} {Nature Communications}\ }\textbf {\bibinfo {volume} {12}},\ \href {https://doi.org/10.1038/s41467-021-22583-5} {10.1038/s41467-021-22583-5} (\bibinfo {year} {2021})\BibitemShut {NoStop}%
\bibitem [{\citenamefont {Santini}\ \emph {et~al.}(2022)\citenamefont {Santini}, \citenamefont {Santoro},\ and\ \citenamefont {Collura}}]{Santini2022}%
  \BibitemOpen
  \bibfield  {author} {\bibinfo {author} {\bibfnamefont {A.}~\bibnamefont {Santini}}, \bibinfo {author} {\bibfnamefont {G.~E.}\ \bibnamefont {Santoro}},\ and\ \bibinfo {author} {\bibfnamefont {M.}~\bibnamefont {Collura}},\ }\bibfield  {title} {\bibinfo {title} {Clean two-dimensional floquet time crystal},\ }\href {https://doi.org/10.1103/PhysRevB.106.134301} {\bibfield  {journal} {\bibinfo  {journal} {Phys. Rev. B}\ }\textbf {\bibinfo {volume} {106}},\ \bibinfo {pages} {134301} (\bibinfo {year} {2022})}\BibitemShut {NoStop}%
\bibitem [{\citenamefont {Collura}\ \emph {et~al.}(2022)\citenamefont {Collura}, \citenamefont {De~Luca}, \citenamefont {Rossini},\ and\ \citenamefont {Lerose}}]{Collura2022}%
  \BibitemOpen
  \bibfield  {author} {\bibinfo {author} {\bibfnamefont {M.}~\bibnamefont {Collura}}, \bibinfo {author} {\bibfnamefont {A.}~\bibnamefont {De~Luca}}, \bibinfo {author} {\bibfnamefont {D.}~\bibnamefont {Rossini}},\ and\ \bibinfo {author} {\bibfnamefont {A.}~\bibnamefont {Lerose}},\ }\bibfield  {title} {\bibinfo {title} {Discrete time-crystalline response stabilized by domain-wall confinement},\ }\href {https://doi.org/10.1103/PhysRevX.12.031037} {\bibfield  {journal} {\bibinfo  {journal} {Phys. Rev. X}\ }\textbf {\bibinfo {volume} {12}},\ \bibinfo {pages} {031037} (\bibinfo {year} {2022})}\BibitemShut {NoStop}%
\bibitem [{\citenamefont {Huang}(2023)}]{Huang2023}%
  \BibitemOpen
  \bibfield  {author} {\bibinfo {author} {\bibfnamefont {B.}~\bibnamefont {Huang}},\ }\bibfield  {title} {\bibinfo {title} {Analytical theory of cat scars with discrete time-crystalline dynamics in floquet systems},\ }\href {https://doi.org/10.1103/PhysRevB.108.104309} {\bibfield  {journal} {\bibinfo  {journal} {Phys. Rev. B}\ }\textbf {\bibinfo {volume} {108}},\ \bibinfo {pages} {104309} (\bibinfo {year} {2023})}\BibitemShut {NoStop}%
\bibitem [{\citenamefont {Shinjo}\ \emph {et~al.}(2024)\citenamefont {Shinjo}, \citenamefont {Seki}, \citenamefont {Shirakawa}, \citenamefont {Sun},\ and\ \citenamefont {Yunoki}}]{Shinjo2024}%
  \BibitemOpen
  \bibfield  {author} {\bibinfo {author} {\bibfnamefont {K.}~\bibnamefont {Shinjo}}, \bibinfo {author} {\bibfnamefont {K.}~\bibnamefont {Seki}}, \bibinfo {author} {\bibfnamefont {T.}~\bibnamefont {Shirakawa}}, \bibinfo {author} {\bibfnamefont {R.-Y.}\ \bibnamefont {Sun}},\ and\ \bibinfo {author} {\bibfnamefont {S.}~\bibnamefont {Yunoki}},\ }\bibfield  {title} {\bibinfo {title} {Unveiling clean two-dimensional discrete time quasicrystals on a digital quantum computer},\ }\href {https://doi.org/10.48550/arXiv.2403.16718} {\bibfield  {journal} {\bibinfo  {journal} {arXiv preprint arXiv:2403.16718}\ } (\bibinfo {year} {2024})}\BibitemShut {NoStop}%
\bibitem [{\citenamefont {Pineda}\ \emph {et~al.}(2014)\citenamefont {Pineda}, \citenamefont {Prosen},\ and\ \citenamefont {Villaseñor}}]{Pineda2014}%
  \BibitemOpen
  \bibfield  {author} {\bibinfo {author} {\bibfnamefont {C.}~\bibnamefont {Pineda}}, \bibinfo {author} {\bibfnamefont {T.}~\bibnamefont {Prosen}},\ and\ \bibinfo {author} {\bibfnamefont {E.}~\bibnamefont {Villaseñor}},\ }\bibfield  {title} {\bibinfo {title} {Two dimensional kicked quantum ising model: dynamical phase transitions},\ }\href {https://doi.org/10.1088/1367-2630/16/12/123044} {\bibfield  {journal} {\bibinfo  {journal} {New Journal of Physics}\ }\textbf {\bibinfo {volume} {16}},\ \bibinfo {pages} {123044} (\bibinfo {year} {2014})}\BibitemShut {NoStop}%
\bibitem [{\citenamefont {Cowtan}\ \emph {et~al.}(2020)\citenamefont {Cowtan}, \citenamefont {Dilkes}, \citenamefont {Duncan}, \citenamefont {Simmons},\ and\ \citenamefont {Sivarajah}}]{Cowtan2020}%
  \BibitemOpen
  \bibfield  {author} {\bibinfo {author} {\bibfnamefont {A.}~\bibnamefont {Cowtan}}, \bibinfo {author} {\bibfnamefont {S.}~\bibnamefont {Dilkes}}, \bibinfo {author} {\bibfnamefont {R.}~\bibnamefont {Duncan}}, \bibinfo {author} {\bibfnamefont {W.}~\bibnamefont {Simmons}},\ and\ \bibinfo {author} {\bibfnamefont {S.}~\bibnamefont {Sivarajah}},\ }\bibfield  {title} {\bibinfo {title} {Phase gadget synthesis for shallow circuits},\ }\href {https://doi.org/10.4204/eptcs.318.13} {\bibfield  {journal} {\bibinfo  {journal} {Electronic Proceedings in Theoretical Computer Science}\ }\textbf {\bibinfo {volume} {318}},\ \bibinfo {pages} {213–228} (\bibinfo {year} {2020})}\BibitemShut {NoStop}%
\bibitem [{\citenamefont {Sur}\ and\ \citenamefont {Sen}(2023)}]{Sur2023}%
  \BibitemOpen
  \bibfield  {author} {\bibinfo {author} {\bibfnamefont {S.}~\bibnamefont {Sur}}\ and\ \bibinfo {author} {\bibfnamefont {D.}~\bibnamefont {Sen}},\ }\bibfield  {title} {\bibinfo {title} {Effects of topological and non-topological edge states on information propagation and scrambling in a floquet spin chain},\ }\href {https://doi.org/10.1088/1361-648X/ad1363} {\bibfield  {journal} {\bibinfo  {journal} {Journal of Physics: Condensed Matter}\ }\textbf {\bibinfo {volume} {36}},\ \bibinfo {pages} {125402} (\bibinfo {year} {2023})}\BibitemShut {NoStop}%
\bibitem [{\citenamefont {Sedlmayr}\ \emph {et~al.}(2023)\citenamefont {Sedlmayr}, \citenamefont {Cheraghi},\ and\ \citenamefont {Sedlmayr}}]{Sedlmayr2023}%
  \BibitemOpen
  \bibfield  {author} {\bibinfo {author} {\bibfnamefont {M.}~\bibnamefont {Sedlmayr}}, \bibinfo {author} {\bibfnamefont {H.}~\bibnamefont {Cheraghi}},\ and\ \bibinfo {author} {\bibfnamefont {N.}~\bibnamefont {Sedlmayr}},\ }\bibfield  {title} {\bibinfo {title} {Information trapping by topologically protected edge states: Scrambling and the butterfly velocity},\ }\href {https://doi.org/10.1103/PhysRevB.108.184303} {\bibfield  {journal} {\bibinfo  {journal} {Phys. Rev. B}\ }\textbf {\bibinfo {volume} {108}},\ \bibinfo {pages} {184303} (\bibinfo {year} {2023})}\BibitemShut {NoStop}%
\bibitem [{\citenamefont {Swingle}\ and\ \citenamefont {Yunger~Halpern}(2018)}]{Swingle2018}%
  \BibitemOpen
  \bibfield  {author} {\bibinfo {author} {\bibfnamefont {B.}~\bibnamefont {Swingle}}\ and\ \bibinfo {author} {\bibfnamefont {N.}~\bibnamefont {Yunger~Halpern}},\ }\bibfield  {title} {\bibinfo {title} {Resilience of scrambling measurements},\ }\href {https://doi.org/10.1103/PhysRevA.97.062113} {\bibfield  {journal} {\bibinfo  {journal} {Phys. Rev. A}\ }\textbf {\bibinfo {volume} {97}},\ \bibinfo {pages} {062113} (\bibinfo {year} {2018})}\BibitemShut {NoStop}%
\bibitem [{\citenamefont {Vovrosh}\ \emph {et~al.}(2021)\citenamefont {Vovrosh}, \citenamefont {Khosla}, \citenamefont {Greenaway}, \citenamefont {Self}, \citenamefont {Kim},\ and\ \citenamefont {Knolle}}]{Vovrosh2021}%
  \BibitemOpen
  \bibfield  {author} {\bibinfo {author} {\bibfnamefont {J.}~\bibnamefont {Vovrosh}}, \bibinfo {author} {\bibfnamefont {K.~E.}\ \bibnamefont {Khosla}}, \bibinfo {author} {\bibfnamefont {S.}~\bibnamefont {Greenaway}}, \bibinfo {author} {\bibfnamefont {C.}~\bibnamefont {Self}}, \bibinfo {author} {\bibfnamefont {M.~S.}\ \bibnamefont {Kim}},\ and\ \bibinfo {author} {\bibfnamefont {J.}~\bibnamefont {Knolle}},\ }\bibfield  {title} {\bibinfo {title} {Simple mitigation of global depolarizing errors in quantum simulations},\ }\href {https://doi.org/10.1103/PhysRevE.104.035309} {\bibfield  {journal} {\bibinfo  {journal} {Phys. Rev. E}\ }\textbf {\bibinfo {volume} {104}},\ \bibinfo {pages} {035309} (\bibinfo {year} {2021})}\BibitemShut {NoStop}%
\bibitem [{\citenamefont {Urbanek}\ \emph {et~al.}(2021)\citenamefont {Urbanek}, \citenamefont {Nachman}, \citenamefont {Pascuzzi}, \citenamefont {He}, \citenamefont {Bauer},\ and\ \citenamefont {de~Jong}}]{Urbanek2021}%
  \BibitemOpen
  \bibfield  {author} {\bibinfo {author} {\bibfnamefont {M.}~\bibnamefont {Urbanek}}, \bibinfo {author} {\bibfnamefont {B.}~\bibnamefont {Nachman}}, \bibinfo {author} {\bibfnamefont {V.~R.}\ \bibnamefont {Pascuzzi}}, \bibinfo {author} {\bibfnamefont {A.}~\bibnamefont {He}}, \bibinfo {author} {\bibfnamefont {C.~W.}\ \bibnamefont {Bauer}},\ and\ \bibinfo {author} {\bibfnamefont {W.~A.}\ \bibnamefont {de~Jong}},\ }\bibfield  {title} {\bibinfo {title} {Mitigating depolarizing noise on quantum computers with noise-estimation circuits},\ }\href {https://doi.org/10.1103/PhysRevLett.127.270502} {\bibfield  {journal} {\bibinfo  {journal} {Phys. Rev. Lett.}\ }\textbf {\bibinfo {volume} {127}},\ \bibinfo {pages} {270502} (\bibinfo {year} {2021})}\BibitemShut {NoStop}%
\bibitem [{\citenamefont {Mi}\ \emph {et~al.}(2021)\citenamefont {Mi}, \citenamefont {Roushan}, \citenamefont {Quintana}, \citenamefont {Mandrà}, \citenamefont {Marshall}, \citenamefont {Neill}, \citenamefont {Arute}, \citenamefont {Arya}, \citenamefont {Atalaya}, \citenamefont {Babbush}, \citenamefont {Bardin}, \citenamefont {Barends}, \citenamefont {Basso}, \citenamefont {Bengtsson}, \citenamefont {Boixo}, \citenamefont {Bourassa}, \citenamefont {Broughton}, \citenamefont {Buckley}, \citenamefont {Buell}, \citenamefont {Burkett}, \citenamefont {Bushnell}, \citenamefont {Chen}, \citenamefont {Chiaro}, \citenamefont {Collins}, \citenamefont {Courtney}, \citenamefont {Demura}, \citenamefont {Derk}, \citenamefont {Dunsworth}, \citenamefont {Eppens}, \citenamefont {Erickson}, \citenamefont {Farhi}, \citenamefont {Fowler}, \citenamefont {Foxen}, \citenamefont {Gidney}, \citenamefont {Giustina}, \citenamefont {Gross}, \citenamefont {Harrigan}, \citenamefont {Harrington}, \citenamefont {Hilton}, \citenamefont
  {Ho}, \citenamefont {Hong}, \citenamefont {Huang}, \citenamefont {Huggins}, \citenamefont {Ioffe}, \citenamefont {Isakov}, \citenamefont {Jeffrey}, \citenamefont {Jiang}, \citenamefont {Jones}, \citenamefont {Kafri}, \citenamefont {Kelly}, \citenamefont {Kim}, \citenamefont {Kitaev}, \citenamefont {Klimov}, \citenamefont {Korotkov}, \citenamefont {Kostritsa}, \citenamefont {Landhuis}, \citenamefont {Laptev}, \citenamefont {Lucero}, \citenamefont {Martin}, \citenamefont {McClean}, \citenamefont {McCourt}, \citenamefont {McEwen}, \citenamefont {Megrant}, \citenamefont {Miao}, \citenamefont {Mohseni}, \citenamefont {Montazeri}, \citenamefont {Mruczkiewicz}, \citenamefont {Mutus}, \citenamefont {Naaman}, \citenamefont {Neeley}, \citenamefont {Newman}, \citenamefont {Niu}, \citenamefont {O’Brien}, \citenamefont {Opremcak}, \citenamefont {Ostby}, \citenamefont {Pato}, \citenamefont {Petukhov}, \citenamefont {Redd}, \citenamefont {Rubin}, \citenamefont {Sank}, \citenamefont {Satzinger}, \citenamefont {Shvarts},
  \citenamefont {Strain}, \citenamefont {Szalay}, \citenamefont {Trevithick}, \citenamefont {Villalonga}, \citenamefont {White}, \citenamefont {Yao}, \citenamefont {Yeh}, \citenamefont {Zalcman}, \citenamefont {Neven}, \citenamefont {Aleiner}, \citenamefont {Kechedzhi}, \citenamefont {Smelyanskiy},\ and\ \citenamefont {Chen}}]{Mi2021}%
  \BibitemOpen
  \bibfield  {author} {\bibinfo {author} {\bibfnamefont {X.}~\bibnamefont {Mi}}, \bibinfo {author} {\bibfnamefont {P.}~\bibnamefont {Roushan}}, \bibinfo {author} {\bibfnamefont {C.}~\bibnamefont {Quintana}}, \bibinfo {author} {\bibfnamefont {S.}~\bibnamefont {Mandrà}}, \bibinfo {author} {\bibfnamefont {J.}~\bibnamefont {Marshall}}, \bibinfo {author} {\bibfnamefont {C.}~\bibnamefont {Neill}}, \bibinfo {author} {\bibfnamefont {F.}~\bibnamefont {Arute}}, \bibinfo {author} {\bibfnamefont {K.}~\bibnamefont {Arya}}, \bibinfo {author} {\bibfnamefont {J.}~\bibnamefont {Atalaya}}, \bibinfo {author} {\bibfnamefont {R.}~\bibnamefont {Babbush}}, \bibinfo {author} {\bibfnamefont {J.~C.}\ \bibnamefont {Bardin}}, \bibinfo {author} {\bibfnamefont {R.}~\bibnamefont {Barends}}, \bibinfo {author} {\bibfnamefont {J.}~\bibnamefont {Basso}}, \bibinfo {author} {\bibfnamefont {A.}~\bibnamefont {Bengtsson}}, \bibinfo {author} {\bibfnamefont {S.}~\bibnamefont {Boixo}}, \bibinfo {author} {\bibfnamefont {A.}~\bibnamefont {Bourassa}},
  \bibinfo {author} {\bibfnamefont {M.}~\bibnamefont {Broughton}}, \bibinfo {author} {\bibfnamefont {B.~B.}\ \bibnamefont {Buckley}}, \bibinfo {author} {\bibfnamefont {D.~A.}\ \bibnamefont {Buell}}, \bibinfo {author} {\bibfnamefont {B.}~\bibnamefont {Burkett}}, \bibinfo {author} {\bibfnamefont {N.}~\bibnamefont {Bushnell}}, \bibinfo {author} {\bibfnamefont {Z.}~\bibnamefont {Chen}}, \bibinfo {author} {\bibfnamefont {B.}~\bibnamefont {Chiaro}}, \bibinfo {author} {\bibfnamefont {R.}~\bibnamefont {Collins}}, \bibinfo {author} {\bibfnamefont {W.}~\bibnamefont {Courtney}}, \bibinfo {author} {\bibfnamefont {S.}~\bibnamefont {Demura}}, \bibinfo {author} {\bibfnamefont {A.~R.}\ \bibnamefont {Derk}}, \bibinfo {author} {\bibfnamefont {A.}~\bibnamefont {Dunsworth}}, \bibinfo {author} {\bibfnamefont {D.}~\bibnamefont {Eppens}}, \bibinfo {author} {\bibfnamefont {C.}~\bibnamefont {Erickson}}, \bibinfo {author} {\bibfnamefont {E.}~\bibnamefont {Farhi}}, \bibinfo {author} {\bibfnamefont {A.~G.}\ \bibnamefont {Fowler}},
  \bibinfo {author} {\bibfnamefont {B.}~\bibnamefont {Foxen}}, \bibinfo {author} {\bibfnamefont {C.}~\bibnamefont {Gidney}}, \bibinfo {author} {\bibfnamefont {M.}~\bibnamefont {Giustina}}, \bibinfo {author} {\bibfnamefont {J.~A.}\ \bibnamefont {Gross}}, \bibinfo {author} {\bibfnamefont {M.~P.}\ \bibnamefont {Harrigan}}, \bibinfo {author} {\bibfnamefont {S.~D.}\ \bibnamefont {Harrington}}, \bibinfo {author} {\bibfnamefont {J.}~\bibnamefont {Hilton}}, \bibinfo {author} {\bibfnamefont {A.}~\bibnamefont {Ho}}, \bibinfo {author} {\bibfnamefont {S.}~\bibnamefont {Hong}}, \bibinfo {author} {\bibfnamefont {T.}~\bibnamefont {Huang}}, \bibinfo {author} {\bibfnamefont {W.~J.}\ \bibnamefont {Huggins}}, \bibinfo {author} {\bibfnamefont {L.~B.}\ \bibnamefont {Ioffe}}, \bibinfo {author} {\bibfnamefont {S.~V.}\ \bibnamefont {Isakov}}, \bibinfo {author} {\bibfnamefont {E.}~\bibnamefont {Jeffrey}}, \bibinfo {author} {\bibfnamefont {Z.}~\bibnamefont {Jiang}}, \bibinfo {author} {\bibfnamefont {C.}~\bibnamefont {Jones}}, \bibinfo
  {author} {\bibfnamefont {D.}~\bibnamefont {Kafri}}, \bibinfo {author} {\bibfnamefont {J.}~\bibnamefont {Kelly}}, \bibinfo {author} {\bibfnamefont {S.}~\bibnamefont {Kim}}, \bibinfo {author} {\bibfnamefont {A.}~\bibnamefont {Kitaev}}, \bibinfo {author} {\bibfnamefont {P.~V.}\ \bibnamefont {Klimov}}, \bibinfo {author} {\bibfnamefont {A.~N.}\ \bibnamefont {Korotkov}}, \bibinfo {author} {\bibfnamefont {F.}~\bibnamefont {Kostritsa}}, \bibinfo {author} {\bibfnamefont {D.}~\bibnamefont {Landhuis}}, \bibinfo {author} {\bibfnamefont {P.}~\bibnamefont {Laptev}}, \bibinfo {author} {\bibfnamefont {E.}~\bibnamefont {Lucero}}, \bibinfo {author} {\bibfnamefont {O.}~\bibnamefont {Martin}}, \bibinfo {author} {\bibfnamefont {J.~R.}\ \bibnamefont {McClean}}, \bibinfo {author} {\bibfnamefont {T.}~\bibnamefont {McCourt}}, \bibinfo {author} {\bibfnamefont {M.}~\bibnamefont {McEwen}}, \bibinfo {author} {\bibfnamefont {A.}~\bibnamefont {Megrant}}, \bibinfo {author} {\bibfnamefont {K.~C.}\ \bibnamefont {Miao}}, \bibinfo {author}
  {\bibfnamefont {M.}~\bibnamefont {Mohseni}}, \bibinfo {author} {\bibfnamefont {S.}~\bibnamefont {Montazeri}}, \bibinfo {author} {\bibfnamefont {W.}~\bibnamefont {Mruczkiewicz}}, \bibinfo {author} {\bibfnamefont {J.}~\bibnamefont {Mutus}}, \bibinfo {author} {\bibfnamefont {O.}~\bibnamefont {Naaman}}, \bibinfo {author} {\bibfnamefont {M.}~\bibnamefont {Neeley}}, \bibinfo {author} {\bibfnamefont {M.}~\bibnamefont {Newman}}, \bibinfo {author} {\bibfnamefont {M.~Y.}\ \bibnamefont {Niu}}, \bibinfo {author} {\bibfnamefont {T.~E.}\ \bibnamefont {O’Brien}}, \bibinfo {author} {\bibfnamefont {A.}~\bibnamefont {Opremcak}}, \bibinfo {author} {\bibfnamefont {E.}~\bibnamefont {Ostby}}, \bibinfo {author} {\bibfnamefont {B.}~\bibnamefont {Pato}}, \bibinfo {author} {\bibfnamefont {A.}~\bibnamefont {Petukhov}}, \bibinfo {author} {\bibfnamefont {N.}~\bibnamefont {Redd}}, \bibinfo {author} {\bibfnamefont {N.~C.}\ \bibnamefont {Rubin}}, \bibinfo {author} {\bibfnamefont {D.}~\bibnamefont {Sank}}, \bibinfo {author}
  {\bibfnamefont {K.~J.}\ \bibnamefont {Satzinger}}, \bibinfo {author} {\bibfnamefont {V.}~\bibnamefont {Shvarts}}, \bibinfo {author} {\bibfnamefont {D.}~\bibnamefont {Strain}}, \bibinfo {author} {\bibfnamefont {M.}~\bibnamefont {Szalay}}, \bibinfo {author} {\bibfnamefont {M.~D.}\ \bibnamefont {Trevithick}}, \bibinfo {author} {\bibfnamefont {B.}~\bibnamefont {Villalonga}}, \bibinfo {author} {\bibfnamefont {T.}~\bibnamefont {White}}, \bibinfo {author} {\bibfnamefont {Z.~J.}\ \bibnamefont {Yao}}, \bibinfo {author} {\bibfnamefont {P.}~\bibnamefont {Yeh}}, \bibinfo {author} {\bibfnamefont {A.}~\bibnamefont {Zalcman}}, \bibinfo {author} {\bibfnamefont {H.}~\bibnamefont {Neven}}, \bibinfo {author} {\bibfnamefont {I.}~\bibnamefont {Aleiner}}, \bibinfo {author} {\bibfnamefont {K.}~\bibnamefont {Kechedzhi}}, \bibinfo {author} {\bibfnamefont {V.}~\bibnamefont {Smelyanskiy}},\ and\ \bibinfo {author} {\bibfnamefont {Y.}~\bibnamefont {Chen}},\ }\bibfield  {title} {\bibinfo {title} {Information scrambling in quantum
  circuits},\ }\href {https://doi.org/10.1126/science.abg5029} {\bibfield  {journal} {\bibinfo  {journal} {Science}\ }\textbf {\bibinfo {volume} {374}},\ \bibinfo {pages} {1479–1483} (\bibinfo {year} {2021})}\BibitemShut {NoStop}%
\bibitem [{\citenamefont {Haegeman}\ \emph {et~al.}(2011)\citenamefont {Haegeman}, \citenamefont {Cirac}, \citenamefont {Osborne}, \citenamefont {Pi\ifmmode~\check{z}\else \v{z}\fi{}orn}, \citenamefont {Verschelde},\ and\ \citenamefont {Verstraete}}]{Haegeman2011}%
  \BibitemOpen
  \bibfield  {author} {\bibinfo {author} {\bibfnamefont {J.}~\bibnamefont {Haegeman}}, \bibinfo {author} {\bibfnamefont {J.~I.}\ \bibnamefont {Cirac}}, \bibinfo {author} {\bibfnamefont {T.~J.}\ \bibnamefont {Osborne}}, \bibinfo {author} {\bibfnamefont {I.}~\bibnamefont {Pi\ifmmode~\check{z}\else \v{z}\fi{}orn}}, \bibinfo {author} {\bibfnamefont {H.}~\bibnamefont {Verschelde}},\ and\ \bibinfo {author} {\bibfnamefont {F.}~\bibnamefont {Verstraete}},\ }\bibfield  {title} {\bibinfo {title} {Time-dependent variational principle for quantum lattices},\ }\href {https://doi.org/10.1103/PhysRevLett.107.070601} {\bibfield  {journal} {\bibinfo  {journal} {Phys. Rev. Lett.}\ }\textbf {\bibinfo {volume} {107}},\ \bibinfo {pages} {070601} (\bibinfo {year} {2011})}\BibitemShut {NoStop}%
\bibitem [{\citenamefont {Fishman}\ \emph {et~al.}(2022)\citenamefont {Fishman}, \citenamefont {White},\ and\ \citenamefont {Stoudenmire}}]{Fishman2022}%
  \BibitemOpen
  \bibfield  {author} {\bibinfo {author} {\bibfnamefont {M.}~\bibnamefont {Fishman}}, \bibinfo {author} {\bibfnamefont {S.~R.}\ \bibnamefont {White}},\ and\ \bibinfo {author} {\bibfnamefont {E.~M.}\ \bibnamefont {Stoudenmire}},\ }\bibfield  {title} {\bibinfo {title} {{The ITensor Software Library for Tensor Network Calculations}},\ }\href {https://doi.org/10.21468/SciPostPhysCodeb.4} {\bibfield  {journal} {\bibinfo  {journal} {SciPost Phys. Codebases}\ ,\ \bibinfo {pages} {4}} (\bibinfo {year} {2022})}\BibitemShut {NoStop}%
\end{thebibliography}%

\onecolumngrid

\foreach \x in {1,...,26}{%
  \clearpage
  \includepdf[pages=\x]{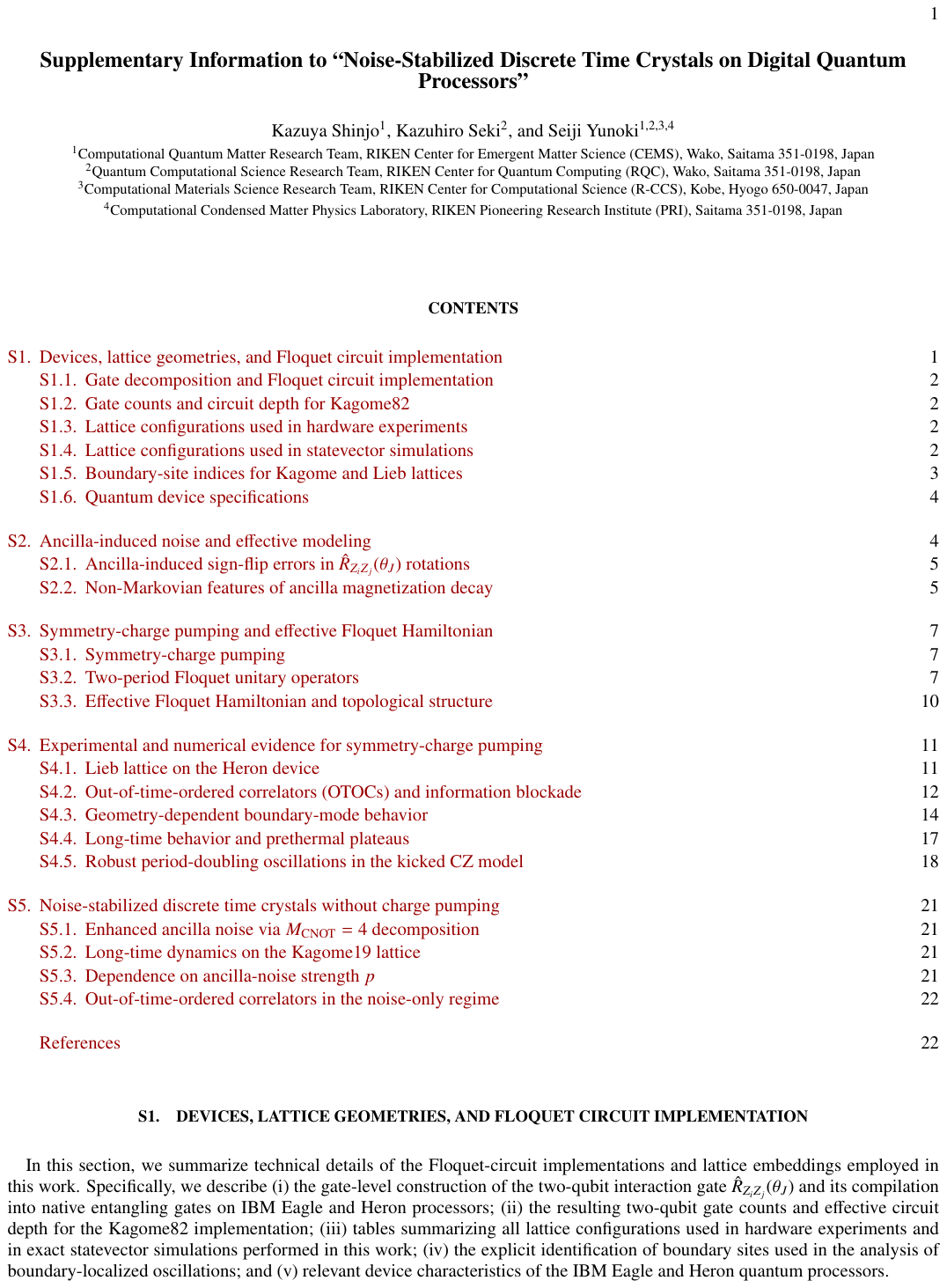}
}

%
%


\end{document}